\documentclass[5p,authoryear]{elsarticle}
\usepackage{amssymb}
\usepackage{amsmath}
\usepackage{url}
\usepackage{caption}
\usepackage{subcaption}
\usepackage{graphicx}
\usepackage{color}

\journal{Astronomy and Computing}

\begin{document}

\newcommand{\code}[1]{\texttt{#1}}

\begin{frontmatter}


\title{Identifying transient and variable sources in radio images}
\author[api,astron]{Antonia Rowlinson}
\ead{b.a.rowlinson@uva.nl}
\author[usyd]{Adam J. Stewart}
\author[astron]{Jess W. Broderick}
\author[seattle]{John D. Swinbank}
\author[api]{Ralph A.M.J. Wijers}
\author[texas]{Dario Carbone}
\author[toronto]{Yvette Cendes}
\author[oxford]{Rob Fender}
\author[wash1,wash2]{Alexander van der Horst}
\author[sa]{Gijs Molenaar}
\author[cwi,dataspex]{Bart Scheers}
\author[oxford]{Tim Staley}
\author[usyd]{Sean Farrell}
\author[LPC2E,srn]{Jean-Mathias Grie{\ss}meier}
\author[techsyd,csiro]{Martin Bell}
\author[taut]{Jochen Eisl{\"o}ffel}
\author[berkley]{Casey J. Law}
\author[astron,api]{Joeri van Leeuwen}
\author[lesia,srn]{Philippe Zarka}
\address[api]{Anton Pannekoek Institute, University of Amsterdam, Postbus 94249, 1090 GE Amsterdam, The Netherlands}
\address[astron]{ASTRON, the Netherlands Institute for Radio Astronomy, Postbus 2, NL-7990 AA Dwingeloo, the Netherlands}
\address[usyd]{Sydney Institute for Astronomy, School of Physics, The University of Sydney, Sydney, NSW, 2006, Australia}
\address[seattle]{University of Washington, Dept. of Astronomy, Box 351580, Seattle, WA 98195}
\address[texas]{Department of Physics and Astronomy, Texas Tech University, Box 1051, Lubbock, TX 79409-1051, USA}
\address[toronto]{Dunlap Institute for Astronomy and Astrophysics, University of Toronto, ON M5S 3H4, Canada}
\address[oxford]{Astrophysics, Department of Physics, University of Oxford, Keble Road, Oxford OX1 3RH, UK}
\address[wash1]{Department of Physics, the George Washington University, 725 21st Street NW, Washington, DC 20052, USA}
\address[wash2]{Astronomy, Physics and Statistics Institute of Sciences (APSIS), 725 21st Street NW, Washington, DC 20052, USA}
\address[sa]{Department of Physics \& Electronics, Rhodes University, Grahamstown, South Africa}
\address[cwi]{CWI, Centrum Wiskunde \& Informatica, PO Box 94079, 1090 GB Amsterdam, The Netherlands}
\address[dataspex]{Dataspex, Science Park 123, 1098 XG Amsterdam, The Netherlands}
\address[LPC2E]{LPC2E - Universit\'{e} d'Orl\'{e}ans /  CNRS, 45071 Orl\'{e}ans cedex 2, France}
\address[srn]{Station de Radioastronomie de Nan\c{c}ay, Observatoire de Paris, PSL Research University, CNRS, Univ. Orl\'{e}ans, OSUC, 18330 Nan\c{c}ay, France}
\address[techsyd]{University of Technology Sydney, 15 Broadway, Ultimo NSW 2007, Australia}
\address[csiro]{CSIRO Astronomy and Space Science, PO Box 76, Epping NSW 1710, Australia}
\address[taut]{Th\"uringer Landessternwarte, Sternwarte 5, D-07778 Tautenburg, Germany}
\address[berkley]{Department of Astronomy and Radio Astronomy Lab, University of California, Berkeley, CA 94720, USA}
\address[lesia]{LESIA, Observatoire de Paris, CNRS, PSL, SU, UPD, SPC, Place J. Janssen, Meudon, France}

\begin{abstract}

With the arrival of a number of wide-field snapshot image-plane radio transient surveys, there will be a huge influx of images in the coming years making it impossible to manually analyse the datasets. Automated pipelines to process the information stored in the images are being developed, such as the LOFAR Transients Pipeline, outputting light curves and various transient parameters. These pipelines have a number of tuneable parameters that  require training to meet the survey requirements. This paper utilises both observed and simulated datasets to demonstrate different machine learning strategies that can be used to train these parameters. We use a simple anomaly detection algorithm and a penalised logistic regression algorithm. The datasets used are from LOFAR observations and we process the data using the LOFAR Transients Pipeline; however the strategies developed are applicable to any light curve datasets at different frequencies and can be adapted to different automated pipelines. These machine learning strategies are publicly available as {\sc Python} tools that can be downloaded and adapted to different datasets (https://github.com/AntoniaR/TraP\_ML\_tools).

\end{abstract}

\begin{keyword}
methods: data analysis \sep methods: statistical \sep radio continuum: general

\end{keyword}

\end{frontmatter}

\section{Introduction}

Transient and variable astronomy is entering an exciting era with vast numbers of images, covering large fields of view, requiring real-time automated processing. In radio astronomy in particular, new instruments including LOFAR \citep[the Low Frequency Array;][]{vanHaarlem:2013}, AARTFAAC  \citep[Amsterdam-ASTRON Radio Transients Facility And Analysis Centre;][]{Prasad:2012}, MWA \citep[Murchison Wide-field Array;][]{Tingay:2013}, LWA1 \citep[Long Wavelength Array Station 1;][]{Ellingson:2013}, ASKAP \citep[Australian Square Kilometre Array Pathfinder;][]{Johnston:2007,Murphy:2013} and MeerKAT \citep{Booth:2009} are exploring unfamiliar frequency and transient duration regimes, opening up the possibility of discovering new types of variable sources. Significant effort is being applied to developing high quality tools that are capable of rapidly and reliably processing images to obtain light curves of all the sources in the fields imaged and to automatically analyse the light curves to find transient or variable sources\footnote{For clarity purposes, we define ``transient sources'' as those which are new sources detected after the first image in a given dataset and ``variable sources'' as those which have variability in their extracted light curves.}. The LOFAR Transients Pipeline \citep[{\sc TraP};][]{Swinbank:2014} is one such tool that is capable of processing images to find transient and variable sources from a wide range of radio telescopes. To date, TraP has been used by LOFAR, AARTFAAC, the Boolardy Engineering Test Array \citep[BETA, the ASKAP test array;][]{Hotan:2014,Hobbs:2016}, the Arcminute Microkelvin Imager \citep[AMI;][]{Zwart:2008,Staley:2013}, the MWA \citep[][]{Rowlinson:2016} and the Jansky Very Large Array Low-band Ionosphere and Transient Experiment \citep[VLITE;][]{Polisensky:2016}.

Once light curves are obtained and analysed by automated pipelines, there are two key challenges: distinguishing transient and variable sources from stable sources and classification of these sources into known source types. A number of existing transient surveys, at a range of wavelengths, have made good progress in determining key diagnostic parameters for identifying transient sources. One of the most powerful diagnostics is the weighted reduced $\chi^2$, referred to as $\eta$ in this paper, of a fit to the light curve by a constant flux density model, which can easily be converted to a probability that the data are drawn from the fitted model \citep[currently used to search for transients at many wavelengths, including X-ray, optical, microwave and radio; e.g.][]{Bannister:2011, Bower:2011, Palanque:2011, Thyagarajan:2011, Hoffman:2012, Shin:2012, Chen:2013, Croft:2013, Mooley:2013, Williams:2013, Bell:2014, Franzen:2014, Bell:2015}. Many surveys utilise a probability threshold to separate the stable sources from the variable sources. For instance, transient sources have a probability of less than 1\% of being a stable source using the $\eta$ from the fit to a stable source, enabling thresholds to be determined by using a False Detection Rate (FDR, the number of false identifications that are expected to be made).

Several other parameters are also used by transient surveys to characterise the variable sources after identification or with an arbitrary threshold to aid in discriminating between variable and stable sources. These parameters can be source specific, such as the pulsar modulation index \citep[e.g.][]{Esamdin:2004}, or target a specific type of variability behaviour, such as periodicity in the light curve. One of the most common additional parameters is the fractional variability \citep[also known as fractional modulation; e.g.][]{Bannister:2011, Bower:2011, Chen:2013, Croft:2013, Mooley:2013, Bell:2014, Franzen:2014, Bell:2015}, which compares the observed scatter in flux densities to the average flux density  of the source. However, variable sources are rarely identified using the information available in these additional parameters; typically they are only used to characterise known sources. In some cases, multiple thresholds are used to discriminate between stable and variable sources but these thresholds are typically chosen arbitrarily based on knowledge of specific datasets. Therefore, there is additional information available to aid in identifying variable sources but currently not used to its full potential.

Over recent years, machine learning strategies have been increasingly applied to astronomical datasets. Machine learning is a form of artificial intelligence in which computer algorithms learn from data, in either a supervised or unsupervised manner, producing models that can then be applied to new data. For instance, one of the common machine learning applications is for classification. The separation of variable sources (ideally into different types) and stable sources using properties of their light curves is a classification problem and, hence, there are a number of machine learning techniques that could be applied to the datasets. Machine learning algorithms exploit various features about each source that can then be utilised in the algorithm; in the context of variable sources these parameters could include $\eta$ and the fractional variability. The algorithm ``learns'' how to separate the different groups of sources using a training dataset and the algorithm can then be applied within automated pipelines to enable rapid classification of new data points.

There are two key types of algorithms in machine learning that can be used: supervised learning and unsupervised learning. There are particular types of unsupervised learning algorithms that would be a good strategy to find unusual sources in the data, in which the algorithm automatically finds new variable sources within a discovery space and classifies sources with similar properties for later identification by astronomers. However, unsupervised techniques, including some powerful unsupervised deep learning algorithms, often require large amounts of data to train in scenarios such as that outlined in this paper. Here, the populations of transient and variable sources are very rare compared to the stable source population and hence a large dataset is needed to ensure a sufficiently large population of the target sources. The required dataset size required for training unsupervised strategies is very difficult to quantify as it depends upon the number of discrete classification groups (the number of which is unknown), having sufficient data in each expected classification group and the distinctness of each group given the training parameters available. When starting a new transient survey searching for rare sources, these sufficiently large training datasets are not available.  Here, as outlined later, we create a labelled dataset with simulated sources making this problem well suited for supervised machine learning strategies. In future surveys, when these issues have been addressed, algorithms such as unsupervised deep learning are likely to prove useful. This is because there will be a large amount of parameters for each unique source and a powerful unsupervised algorithm may identify previously undiscovered transient or variable behaviour of the sources.

Alternatively, supervised machine learning strategies can be trained to find and classify different types of sources. These strategies use training datasets that contain pre-classified sources, which can either be from manual identification of sources in real data or from simulated datasets. For instance, supervised machine learning strategies can be used to reliably discriminate between new sources and imaging artefacts \citep[e.g.][]{Bloom:2012, Brink:2013}. Additionally, a supervised machine learning algorithm, known as a Random Forest \citep{Breiman:2001}, has been trained to identify different types of optical sources using data from the Palomar Transients Factory \citep[PTF;][]{Bloom:2012}. A Random Forest technique has also been applied to 2XMM and 3XMM data \citep[the second and third releases of the X-ray serendipitous source catalogue from the {\it XMM-Newton} satellite;][]{Jansen:2001,Watson:2009,Rosen:2016} to enable optimal identification of specific types of sources \citep{Lo:2014a,Lo:2014b,Farrell:2014} and to the classification of fast radio bursts \citep[e.g][]{Wagstaff:2016}. The Random Forest algorithm is also being investigated as a classification tool within the VAST (Variable and Slow Transient) survey which will be conducted using ASKAP \citep{Murphy:2013}. However, supervised learning techniques are really good at identifying objects that fit into known classes, but not good at identifying objects that belong to new classes. Unfortunately, when exploring a completely new parameter space, this information is often unknown or speculative. In this work, we instead focus on the simpler classification of sources into the categories of transient, variable or stable, which can be completed without knowing the behaviour of specific transient sources. Alternatively, some studies are now looking at unsupervised machine learning techniques such as unsupervised deep learning algorithms \citep[e.g.][]{Connor:2018}.

In this paper, we use the {\sc TraP} to analyse variability of sources within large datasets from LOFAR, both observational and simulated. The transient surveys conducted by LOFAR are searching a relatively unexplored parameter space in radio astronomy: wide fields of view (enabling detection of the rarest variable sources) at low frequencies and on a range of time scales. Hence, this new parameter space may contain previously unidentified categories of variable sources, so we do not have the labelled datasets required to train machine learning strategies such as Random Forest algorithms to find these new sources. Instead, we utilise various variability parameters, output by {\sc TraP}, combined with machine learning techniques, to determine thresholds and algorithms that can optimally separate variable sources from stable sources. We create a simple labelled dataset that can be used to train and test the algorithms used. In Section 2, we describe the LOFAR datasets used in developing these new techniques. Section 3 presents analyses of the variability parameters output from {\sc TraP} for the datasets. The implementation of machine learning strategies is explained in Section 4, including a simple anomaly detection algorithm, as proposed by \cite{Denning:1987}, that outputs thresholds suitable for use in {\sc TraP} and a penalised logistic regression algorithm \citep[e.g.][]{Darroch:1972} that uses multiple parameters to categorise datasets. Finally, in Section 5, we discuss future improvements to these strategies.

\section{Training datasets}

\subsection{LOFAR Observations}

In order to design and test new methods of identifying transient and variable sources, we require a large dataset containing many different sources. The LOFAR Transients Key Science Project has conducted a large survey, the Radio Sky Monitor (RSM), the first large field-of-view and systematic transient survey conducted by LOFAR \citep{Fender:2008}. The RSM dataset is a zenith monitoring survey, with the observations used in this paper occurring on time scales ranging from monthly to approximately 1 year from February 2013 to January 2014. One snapshot of the RSM comprises images from 6 overlapping pointing directions and, over the full 24 hour observation, surveys the zenith strip above LOFAR which is centred on a latitude of 53$^\circ$. In this analysis we exclude fields near the bright ``A-Team'' sources (the brightest sources in the Northern Hemisphere, such as Cygnus A and Cassiopeia A) and fields covering the Galactic plane as these fields can be prone to imaging artifacts\footnote{Note, with careful manual manipulation it is possible to use these observations \citep[e.g. ][]{Broderick:2016}}. Each pointing direction is observed twice in consecutive 11 minute snapshot observations, which we combine into a single 22 minute image, and the full observation is repeated 7 times. The first 6 observations are separated by $\sim$1 month with the final observation being $\sim$6--12 months afterwards. In the dataset used for this analysis, each field is observed in 4 frequency bands (124, 149, 156 and 185 MHz) with a bandwidth of $\sim$2 MHz. The images were made using projected baseline lengths up to 3 k$\lambda$ (corresponding to 6 km at 150 MHz) and the primary beam full width half maximum is 3.8 degrees at 150 MHz. The full dataset utilised in this analysis consists of 5120 images, with the analysis area from a single full observation covering $\sim$970 square degrees in each of 4 frequency bands, with an average rms (root mean square) noise of $23^{+5}_{-3}$ mJy beam$^{-1}$. The data are calibrated and imaged using the strategy outlined in \cite{Broderick:2016,Carbone:2016,Stewart:2016}. The typical resolution is 60 arcsec $\times$ 30 arcsec and the images are not confusion noise limited.

\subsection{LOFAR Simulations}

\begin{figure}
\centering
\includegraphics[width=0.48\textwidth]{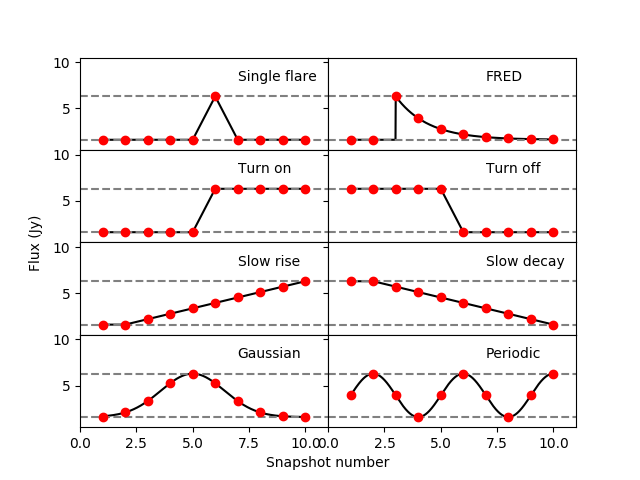}
\caption{Examples of the different transient light curves simulated. The solid line represents the underlying transient light curve and the data points are from an example dataset. The dashed lines show the quiescent flux density  and the peak flux density  (1.6 and 6.3 Jy respectively) used in this example.}
\label{fig:sim_trans}
\end{figure}

To date, we have only identified a small number of transient and variable sources observed at low radio frequencies \citep[e.g.][]{Stewart:2016,Broderick:2016,Murphy:2017}. Therefore, to test the methods presented in this paper, we needed to create simulated datasets containing variable sources. We have produced a large number of simulated LOFAR datasets containing a range of variable sources using the following method:

\begin{enumerate}
    \item To create the simulations, we have chosen a single 11 minute HBA target observation constituting 10 sub-bands centred on 156 MHz, where 1 sub-band has a bandwidth of 195 kHz. The observed data were deleted from the individual sub-band measurement sets leaving the LOFAR structure and required metadata. Random, unique Gaussian noise is inserted into each individual baseline in the measurement set, differentiating between different antenna types and station configurations, with a mean value equal to the system equivalent flux density (SEFD) calculated according to the method described in \cite{Nijboer:2013}. 
    \item We created a range of simulated transient and variable sources, with each source having 10 time steps. The sources range in peak flux density  (the maximum flux density  the variable source reaches) from 0.2 to 50 Jy, with 10 different values in this range separated equally in log space. Each source has a quiescent source of flux density  0 Jy up to the peak flux density  value (in the same bins as the maximum flux density ),  giving a total of 55 different combinations for each source type. Eight different variability types are simulated and illustrated in Figure \ref{fig:sim_trans}: single flare, turn on, turn off, slow rise, slow fall, fast rise exponential decay (FRED), Gaussian and periodic. These variability types are chosen to represent a range of behaviour that we might expect to observe. We note that the snapshots, used in this paper and often for other datasets, are irregularly spaced. Thus, the simulations could be associated with a range of different transient phenomena in the time domain and would not necessarily show the same light curve shape when plotted against the time.
    \item For each simulated dataset, containing 1 variable source, we created a random sky model containing 50 point sources. The sources have random positions within the image and a random flux density  within the range 0--100 Jy sampled from a power-law distribution of flux densities with a slope of -1.
    \item Each skymodel was inserted, using the LOFAR calibration tools \citep{Loose:2008}, into the 10 sub-bands (measurement sets) containing Gaussian noise and we conducted a calibration step to prepare the data for imaging. The 10 sub-bands were combined into a single measurement set which is centred on 156 MHz and identical to band 3 from the RSM dataset.
    \item Finally, we imaged the resulting dataset using {\sc AWImager} \citep{Tasse:2013} and cleaned each image until the observed rms in the image was $\sim$20 mJy (comparable to the rms in the LOFAR observations described in Section 2.1). The image metadata was edited to create the 10 consecutive snapshot time series for inputting into the {\sc TraP}.
\end{enumerate}

The simulations comprise 440 simulated fields with 10 snapshot observations per field and one simulated transient per field, giving a total dataset of 4400 images.

\subsection{Hybrid Training Dataset}

The simulations produce the variable sources required for testing the methods in the paper; however the simulations do not reproduce a number of factors that can introduce additional variability within the observed dataset. These factors can include instrumental variability, radio frequency interference, correlated noise in the images, ionospheric effects, imaging artefacts and systematic calibration errors (e.g. elevation dependencies). Therefore, in observed data, even the stable sources exhibit low level variability, leading to shifts and skews in the population distributions of the variability parameters, and the simulations are significantly underestimating this contribution. The variable sources will also have this additional low level variability, but their total variability statistics are dominated by the simulated variability.

When training methods to automatically distinguish between the stable and variable sources in new observed datasets, it is important that the stable sources are representative of the final stable source population and the simulations alone cannot do this. Therefore, for training the machine learning algorithms described in Section 4, we produce a hybrid training dataset where the stable sources are represented by all the sources observed in the LOFAR observations described in Section 2.1 and the variable sources are drawn from the simulations described in Section 2.2. This hybrid dataset contains both known variable sources and a population of sources which are representative of the final observed population. We note that the simulated variable sources are not affected by the external factors outlined in the previous paragraph, such as instrumental and ionospheric noise, and this will lead to a reduction in value for their variability parameters. However, as these variability parameters are dominated by the artificial variability inserted for the simulated variable sources, this reduction affect is minimal.

However, this dataset has a significant caveat as we are assuming that the observed population of sources are not variable on the time scales used in the survey. If there is a real population of sources exhibiting low variability, this can bias the results as the classifiers are being trained to treat this population as stable. This caveat is negligible if only a small number of sources are intrinsically variable, as the machine learning methods used model the populations and are not significantly affected by outliers to the distributions. As a consequence, with the hybrid dataset, we cannot train the algorithms to identify a population of real sources with low level variability but it will be able to identify the outliers to the distribution. In Section 5.1, we discuss future methods to mitigate this caveat. 

\section{Automated processing using {\sc TraP}}

We use the LOFAR {\sc TraP} to conduct the following tests: image level quality control, source extraction, source association and transient search \citep[as described in ][]{Swinbank:2014}.

\subsection{{\sc TraP} settings}

\begin{figure}
\centering
\includegraphics[width=0.45\textwidth]{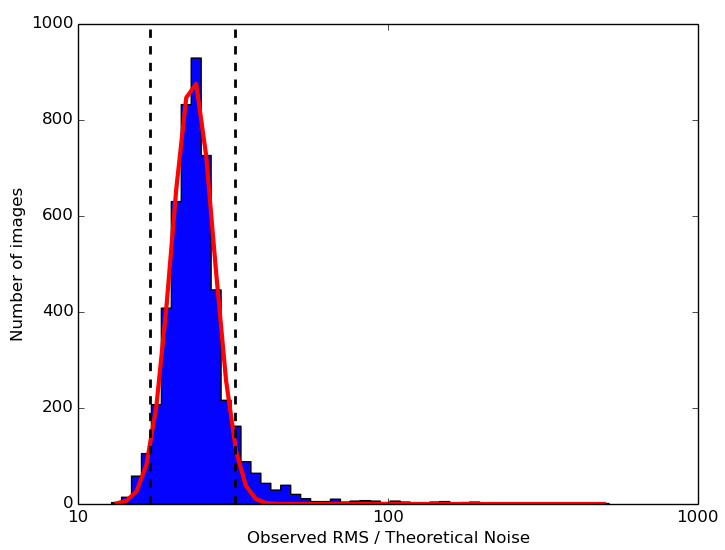}
\caption{The rms noise is automatically measured in each image in the dataset and compared to the theoretical noise expected in the image. Here we plot a histogram of these ratios and they follow a Gaussian distribution in logarithmic space. A Gaussian distribution is fitted to the data (red line) and all images which deviate more than 2$\sigma$ from the observed distribution (shown by the dashed black lines) are excluded from the dataset. Large deviations from this distribution typically signify calibration errors or high rms noise datasets.}
\label{fig:QC}
\end{figure}

The {\sc TraP} quality control settings are applied to the observed RSM dataset (all simulated images are of good quality by design). In Figure \ref{fig:QC}, we plot a histogram of the ratio between the observed rms noise and the theoretical rms noise for all of the images in the dataset. This histogram is fitted with a Gaussian distribution and we reject all images with an rms noise that deviates by more than 2$\sigma$ from the mean, i.e. $>32.1\times$ or $<17.1\times$ the expected theoretical thermal noise, as these images typically have high RFI or calibration errors giving flux density scale errors. We note that we do not achieve the theoretical noise in these images as we do not conduct self calibration or calculate direction dependent effects. Additionally, we reject images where the restoring beam ellipticity ($\frac{B_{\rm maj}}{B_{\rm min}}$, where $B_{\rm maj}$ is the size of the major beam axis and $B_{\rm min}$ is the size of the minor beam axis) is in excess of 1.42; these images are likely to be of low quality as we expect the restoring beam shape to be close to circular because we are conducting a zenith survey. These settings were chosen using an automated quality control diagnostics script that removes images with parameters $>2\sigma$ from the typical distribution of those parameters within the full dataset\footnote{The automated quality control script is available here: https://github.com/transientskp/scripts/tree/master/TraP$\_$QC$\_$diagnostics}. After these quality control settings are applied, we have a dataset comprising 81\% of the images from the original dataset (4017 images).

For source extraction, we use an 8$\sigma$ detection threshold, a 3$\sigma$ analysis threshold and constrain the Gaussian shape fit parameters for all of the sources to be identical to the shape of the restoring beam (assuming all candidate variable sources are point sources)\footnote{We note that assuming all sources are point sources will cause the flux of extended sources to be underestimated \citep{Carbone:2018}. However, this is not an issue for this study as extended sources are not expected to vary on significant time scales and the fitting strategy will ensure that they will be output as stable sources.}. We extract all sources within a radius of 1.3 degrees from the image centre (roughly equivalent to the primary beam FWHM at the highest observing frequency, giving a search area of $\sim$5.4 square degrees per image). Otherwise, we use standard settings for source finding within {\sc TraP} as described in \cite{Swinbank:2014} which utilises the {\sc PySE} source finder \citep{Carbone:2018}. These source extraction parameters resulted in 5224 unique sources in the RSM dataset and 440 variable and transient sources in the simulated datasets.

{\sc TraP} calculates two different parameters for each light curve to quantify the variability of sources. These parameters are the weighted reduced $\chi^2$ ($\eta$, typical to many transient surveys as described in Section 1) and the variability index ($V$; equivalent to the fractional variability used by other surveys); these parameters are consistent with methods used in \cite{Bannister:2011, Bower:2011, Palanque:2011, Thyagarajan:2011, Hoffman:2012, Shin:2012, Chen:2013, Croft:2013, Mooley:2013, Williams:2013, Bell:2014, Franzen:2014}. Additionally, all new sources which were not detected in previous images (when they are expected to be detected by comparison to the detection thresholds in the previous images), are inserted into a transient list. These settings are explained in depth in \cite{Swinbank:2014}.

In this Section, we focus on the transient parameters $\eta_{\nu}$ and $V_{\nu}$, defined as: 
\begin{eqnarray}
\eta_\nu = \frac{N}{N-1}\Biggl(\overline{w_\nu\,{I_\nu}^2}-\frac{\overline{w_\nu\,I_\nu}^2}{\overline{w_\nu}}\Biggr) \label{eqn:eta_nu} \\
V_\nu =\frac{1}{\overline{I_\nu}}\sqrt{\frac{N}{N-1}(\overline{{I_\nu}^2}-\overline{I_\nu}^2)}, \label{eqn:V_nu} 
\end{eqnarray}
where $N$ is the number of data points, $I_{\nu,i}$ is the flux density of a data point at frequency $\nu$, $\overline{w_\nu\,{I_\nu}^2}$ is the average of the weights multiplied by the squared fluxes of all data points for a source, $\overline{w_\nu\,{I_\nu}}$ is the average of the weights multiplied by the fluxes for a source, $\overline{I_{\nu}}$ is the average flux density,  $\overline{I_{\nu}^2}$ is the average squared flux density of a source, $\overline{w_{\nu}} = \frac{1}{N} \sum_{i=0}^N w_{\nu,i} \equiv \frac{1}{N} \sum_{i=0}^N \frac{1}{\sigma_{\nu,i}^2}$,
and $\sigma_{\nu,i}$ is the error on the $i$-th flux density  measurement at frequency $\nu$. In Equation \ref{eqn:eta_nu}, we define the reduced weighted $\chi^2$ using the aggregate quantities as used by {\sc TraP} for each time step \citep[the full derivation of this from the standard reduced weighted $\chi^2$ Equation is given in Section 6.4.4 of ][]{Swinbank:2014}. {\sc TraP} calculates and stores each of these parameters at every frequency and time-step for every source in the dataset. In this analysis, we use the value of these parameters after the full dataset has been processed but the methods presented here can also be applied after each time-step. We note that $\eta$ is roughly proportional to the signal to noise ratio (SNR) squared (where $\omega I^2 = \frac{I^2}{\sigma^2} = SNR^2$) so sources with greater SNR will have a larger value for $\eta$ than lower SNR sources.

The two datasets were processed by {\sc TraP} using these settings and we extract the following parameters for all sources from the database: maximum flux density  that the source attains, the ratio between the maximum flux density  and the average flux density, $\eta_{\nu}$ and $V_{\nu}$. We note that, in the RSM dataset, sources may be detected multiple times in a single time step at a single frequency due to overlapping pointing directions. These multiple detections will be included in Equations \ref{eqn:eta_nu} and \ref{eqn:V_nu}, so can affect the value of these variability parameters. To mitigate this effect, we impose a small source extraction radius to minimise the number of sources in the overlapping regions but maximise the usable image area. We estimate that this affects $\sim$8\% of sources in the RSM dataset and these sources may have multiple catalogue entries for each time step.

\subsection{Results - RSM dataset}

\begin{figure*}
\centering
\includegraphics[width=1.\textwidth]{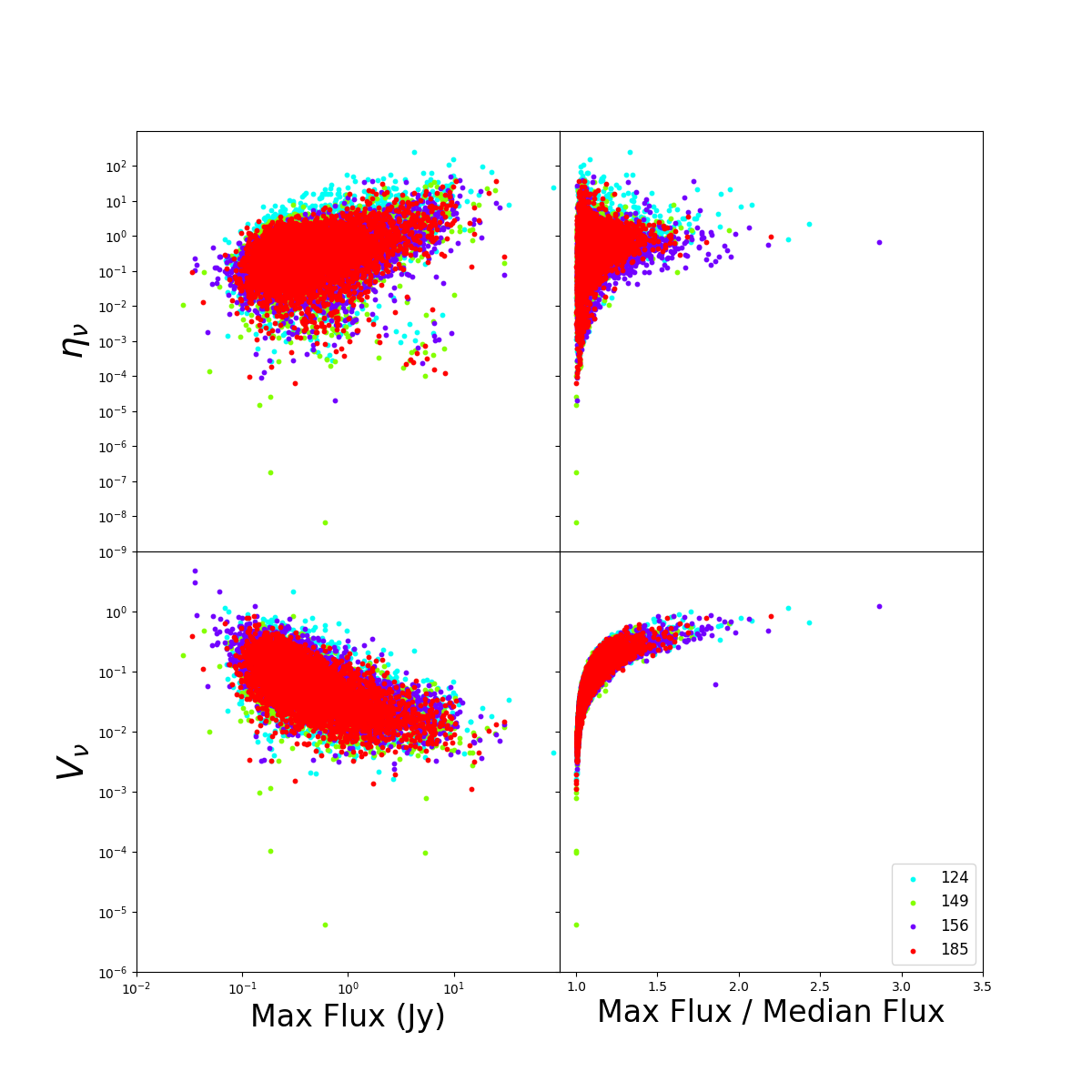}
\caption{The transient parameters for all the sources observed in the RSM dataset. Top row: $\eta_{\nu}$ values. Bottom row: $V_{\nu}$ values. Left column: The maximum flux density of the source. Right column: The maximum ratio between the observed flux density and average flux density for each source. All the sources are colour coded by the observing frequency in MHz.}
\label{fig:rsm_trans_params2}
\end{figure*}

\begin{figure*}
\centering
\includegraphics[width=1.\textwidth]{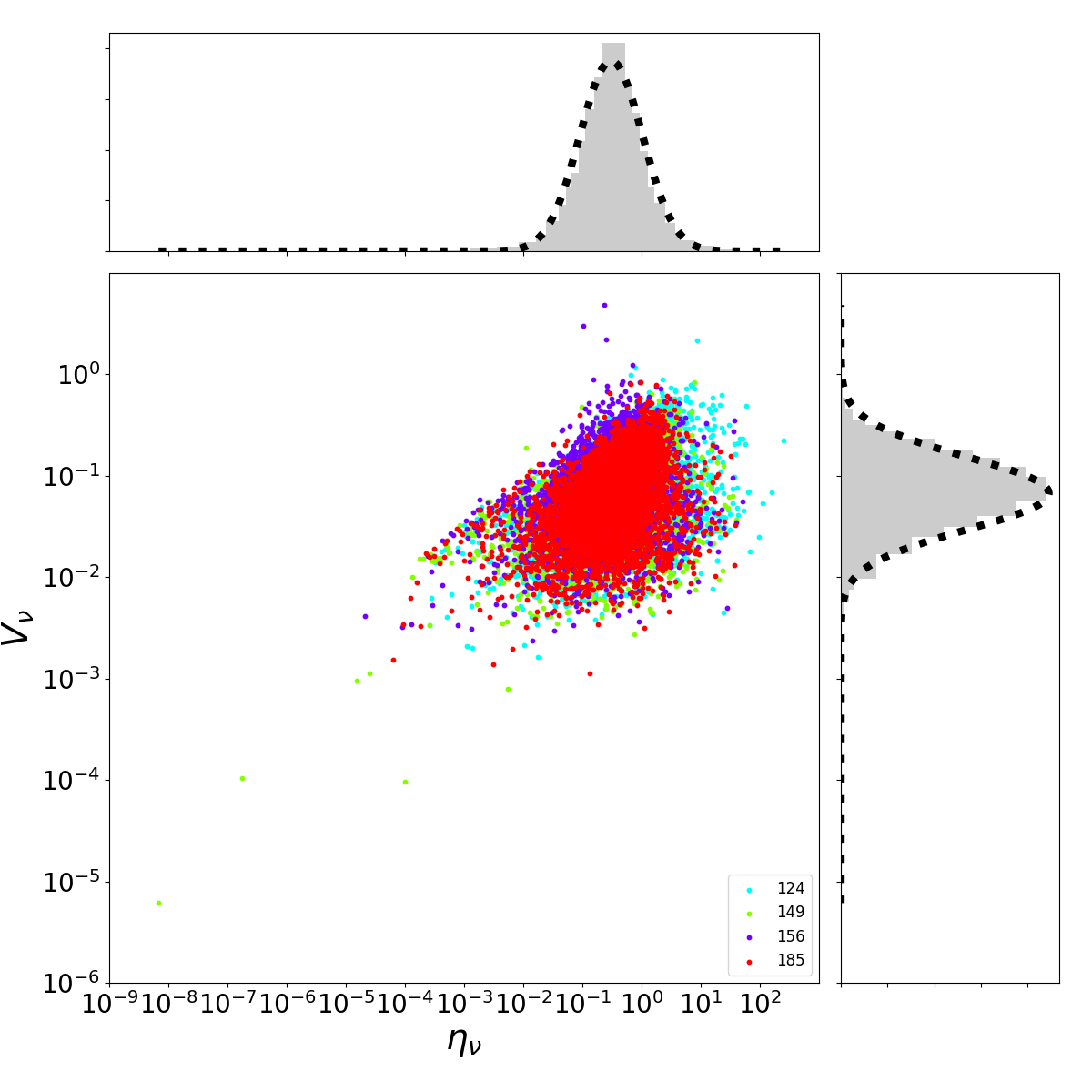}
\caption{The transient parameters for all the sources observed in the RSM dataset. The $\eta_{\nu}$ and $V_{\nu}$ values for all the sources are colour coded by the observing frequency in MHz. The histograms, with Bayesian block binning \citep{Vanderplas:2012,Scargle:2013}, for both $\eta_{\nu}$ and $V_{\nu}$ are well fitted in log space by Gaussian distributions.}
\label{fig:rsm_trans_params1}
\end{figure*}

In Figure \ref{fig:rsm_trans_params2} we introduce a diagnostic plot which illustrates the different parameters for all the observed sources, colour-coded by the different observing frequencies in the RSM dataset. These observing frequencies show similar behaviour, although there is a slightly increased scatter for sources observed at 124 MHz likely caused by a decrease in sensitivity at this frequency relative to the other observing frequencies. From this figure we can draw the following conclusions:
\begin{itemize}
    \item Sources with a high $\eta_{\nu}$ are typically those with a high maximum flux density (as expected from equation \ref{eqn:eta_nu}) but with a low ratio between the maximum flux density  and the average flux density   of the source (and hence a low $V_{\nu}$ but with a large scatter on these values). These are sources where the statistical errors on the flux density measurements are very low relative to the level of the flux density, giving high values of $\eta_{\nu}$ as it is weighted by the statistical errors. However, there are systematic errors in the flux density  values (typically up to 10\% in the RSM dataset), which are not currently being accounted for within {\sc TraP}, and they exceed the statistical errors for these sources leading to a larger value of $\eta_\nu$.
    \item Sources with a high $V_{\nu}$ tend to have very high ratios, between their maximum flux density  and average flux, but very low flux densities (with large uncertainties, and hence a low $\eta_{\nu}$). These sources are typically close to the detection threshold of the images so they are highly sensitive to changes in the structure and amplitude of the image noise which can raise the value of $V_\nu$.
\end{itemize}
These findings are consistent with those determined using comparable parameters from other surveys \cite[e.g.][]{Croft:2013, Mooley:2013}.

By plotting histograms of $\eta_{\nu}$ and $V_{\nu}$, as shown in our second diagnostic plot Figure \ref{fig:rsm_trans_params1}, it is clear they both follow an approximately Gaussian distribution in logarithmic space. As the majority of sources in the field are not expected to be significantly variable, it is therefore reasonable to assume that $\eta_{\nu}$ and $V_{\nu}$ for all stable sources will follow a Gaussian distribution as fitted to the observed data. Any sources whose variability parameters are anomalously large in comparison to these distributions are expected to be variable sources. Therefore, we can identify variable sources using thresholds for $\eta_{\nu}$ and $V_{\nu}$ given by a multiple of $\sigma$ (where $\sigma$ is the standard deviation of the fitted Gaussian distributions).  However, it is not obvious what the optimum thresholds should be to reliably identify variable sources. 

\subsection{Results - Simulated datasets}

In the previous section, we assumed that the majority of the sources detected in the RSM dataset are not variable sources and demonstrated that their transient parameters follow an approximately Gaussian distribution. Variable sources are expected to be those where the transient parameters are anomalous compared to those of the rest of the dataset. However, it is important to test this expectation by identifying the typical parameters for variable sources. The simulated datasets described in Section 2.2 are designed to explore the transient parameter space to quantify which variables are detectable by {\sc TraP}. After running the simulated datasets through {\sc TraP}, we are able to separate the simulated variable sources from the stable sources. In this analysis, we have also removed the sources which are not detected in the first image of the simulation as these trigger the ``new source'' detection strategy within {\sc TraP} (described further in Section 3.3.1). These are the simulated sources where the quiescent flux density is below the detection threshold. This includes all types of sources simulated apart from the ``turn-off'' and ``slow fall'' sources (see Figure \ref{fig:sim_trans}).

\begin{figure}
\centering
\includegraphics[width=.5\textwidth]{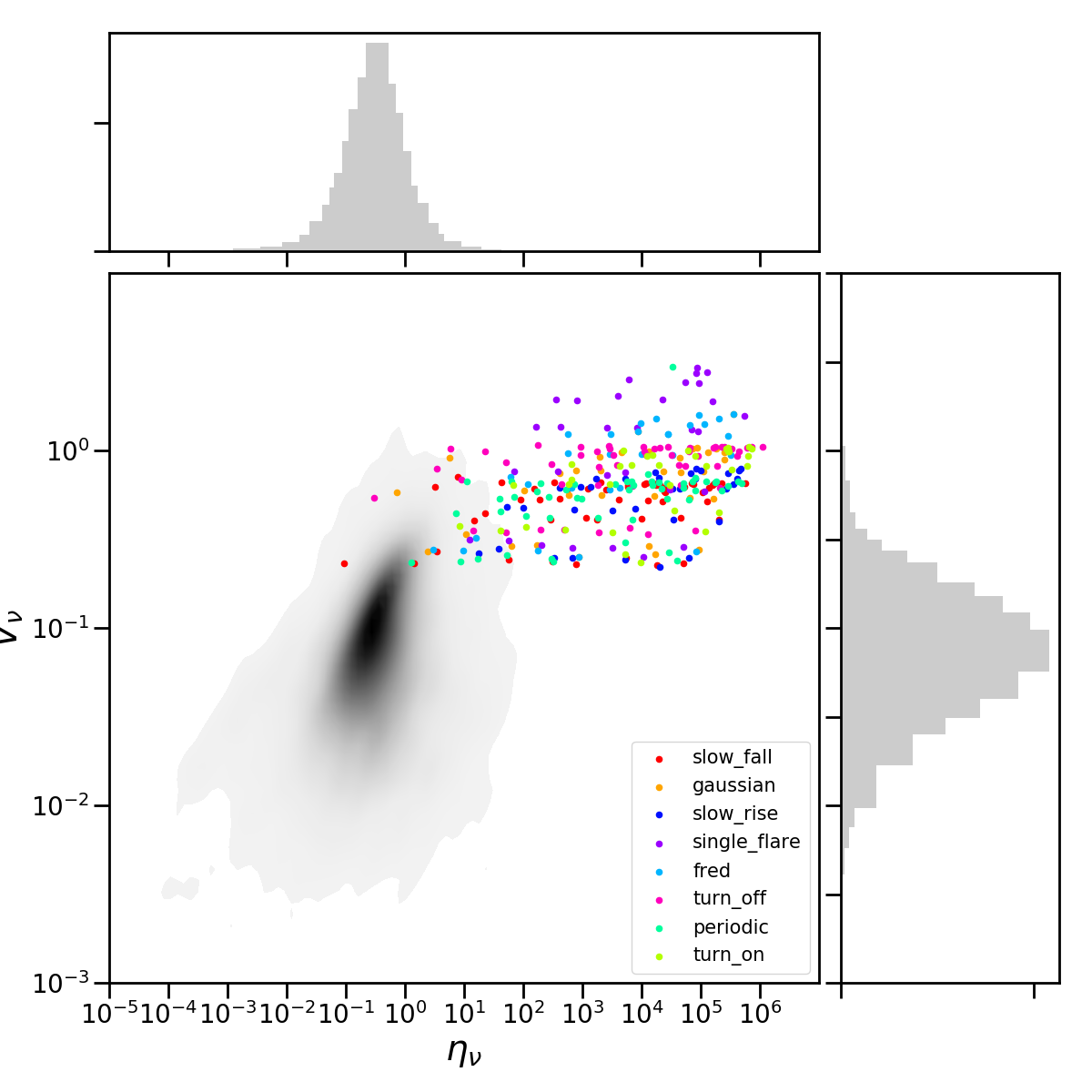}
\caption{This figure, in the same format as Figure \ref{fig:rsm_trans_params1}, shows the transient parameters for all the sources observed in the hybrid dataset (colour scheme in the legend). The grey data are the RSM dataset presented in Figure \ref{fig:rsm_trans_params1} and are assumed to be stable sources. The histograms do not include the simulated transient sources.}
\label{fig:sim_trans_params1}
\end{figure}

\begin{figure}
\centering
\includegraphics[width=.5\textwidth]{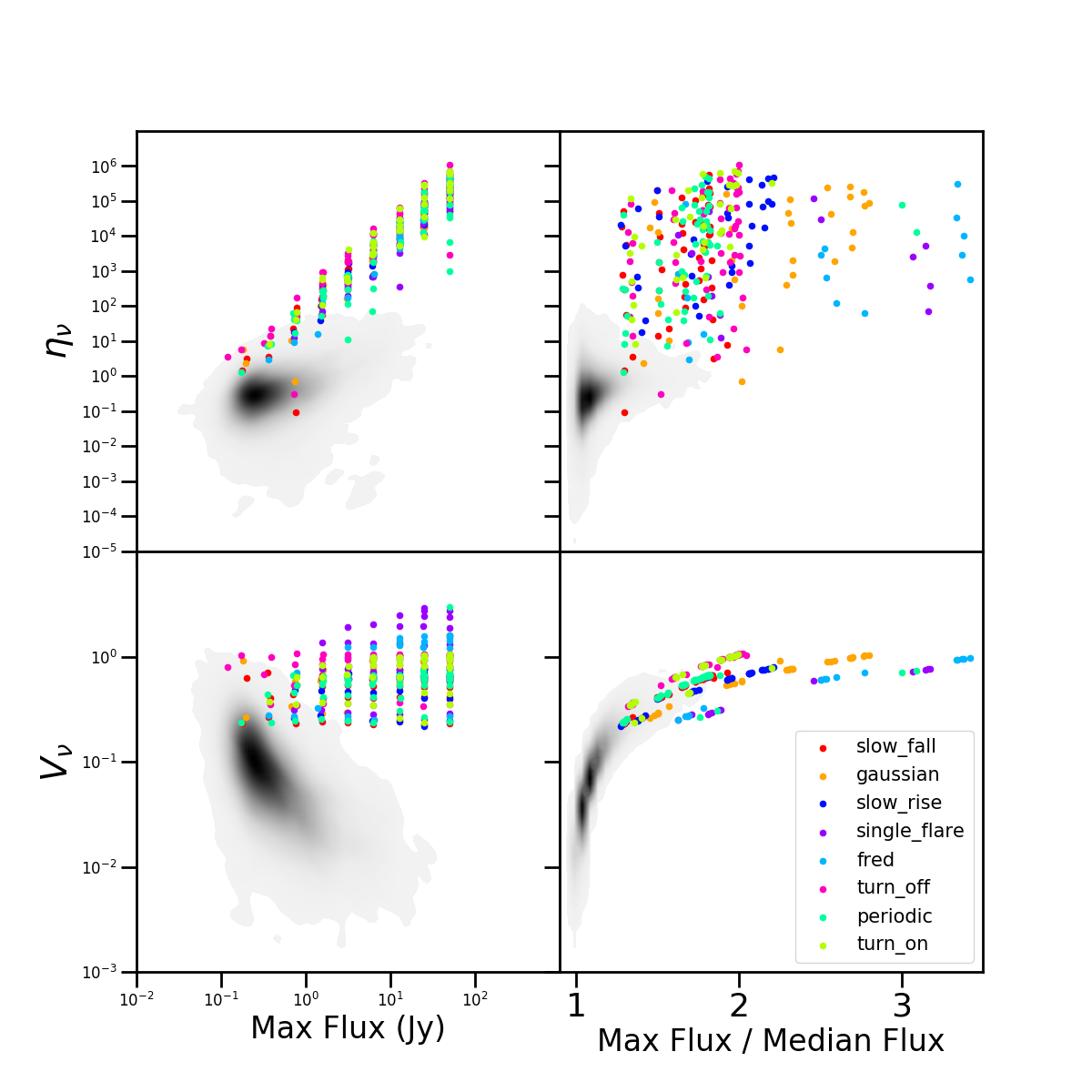}
\caption{As in Figure \ref{fig:rsm_trans_params2}, the transient parameters are plotted for all the sources observed in the simulated datasets (colour scheme in legend and as in Figure \ref{fig:sim_trans_params1}).}
\label{fig:sim_trans_params2}
\end{figure}

The simulated stable sources follow a Gaussian distribution, as seen previously for the RSM dataset; however the stable sources have significantly lower values of $\eta_{\nu}$ and $V_{\nu}$ than the values observed in the RSM dataset. This is because the observed data have additional scatter in the flux densities due to observational effects such as the ionospheric conditions, calibration uncertainties and radio frequency interference which are not accounted for in the simulations. Therefore, we use the hybrid dataset described in Section 2.3, where the RSM data are labelled as stable sources (n.b. this may introduce a bias, as described in Section 2.3).

Figures \ref{fig:sim_trans_params1} and \ref{fig:sim_trans_params2} show the variability properties of the simulated datasets with a clear separation between the variable and stable sources (colour-coded by the type of variable source with the stable sources being taken from the RSM dataset are shown in grey). The ``single flare'' variable sources frequently have higher values for $V_{\nu}$; this is unsurprising as their average flux densities will be significantly lower than the average flux density of the other variable source types. Therefore, the ratio between the maximum flux density and the average flux density can also be much higher than the other sources (as shown in Figure \ref{fig:sim_trans_params2}) leading to the higher values of $V_{\nu}$. This argument is somewhat true for the ``FRED'' (fast rise exponential decay) light curves, as the average flux density is also lower. All the simulated variable sources have variability parameters that typically lie in the same regions of Figures \ref{fig:sim_trans_params1} and \ref{fig:sim_trans_params2} and these are indeed anomalous to the observed Gaussian distributions of $\eta_{\nu}$ and $V_{\nu}$ for sources from the RSM dataset. As the simulated variable sources share the same parameter space, it is reasonable to assume that ``real'' variable sources will also be clearly anomalous to the stable sources (if of sufficient SNR). There are a small number of simulated variable sources that are difficult to distinguish from the stable sources in Figure \ref{fig:sim_trans_params1} and, from Figure \ref{fig:sim_trans_params2}, it is clear that these sources have a ratio between their maximum flux density and average flux density of less than 2 and overlap with the stable sources.

Typically in transient surveys, variable sources are identified using a probability calculated using only $\eta_{\nu}$ and a threshold is determined using a FDR. However, it is clear from Figure \ref{fig:sim_trans_params1} that using a threshold based on the $\eta_{\nu}$ values will either lead to a large number of false detections or fail to identify a large number of transients; however we note that using a threshold on only $\eta_{\nu}$ does produce better results than using a threshold on just the $V_{\nu}$ values. In order to unambiguously identify a large number of transients, we could instead use a combination of thresholds on both the $\eta_{\nu}$ and $V_{\nu}$ values to take full advantage of the separation of variable sources and stable sources.

\subsection{Identification of transient sources}

Sources which appear during the observation are inserted into the database with zeroes for their variability parameters, and hence it may take a number of observations to determine that they are variable sources. This is particularly problematic for sources which are at a roughly constant flux density following their first detection (e.g. the sources which ``turn on'' in the simulated datasets). To aid in rapid identification of these sources, {\sc TraP} instead determines if each new source should be detected in the previous best image (defined as being the image with the lowest rms value previously observed) using the detection threshold plus a small margin. 

The rms within the images is expected to vary significantly; for example the rms is higher in the outer parts of the image where the primary beam is less sensitive and it can be higher in regions containing the side-lobes of bright sources. When the {\sc PySE} source finder runs, it outputs the gridded rms map used for source extraction. {\sc TraP} then records the maximum and minimum rms from within the source extraction region utilised. 

New sources that should be detected in the worst part of the previously best image if they were constant sources are identified as transients and those which would only be detected if in the best part of the image are labelled as candidate transients. For further explanation of this, refer to \cite{Swinbank:2014} Section 5.1.1 and Equations 38 \& 39. To enable fine-tuning to prevent false detections, the margin that is added to the detection threshold is a user defined input although it is not clear what value should be utilised for this margin. This margin is referred to as the new source sigma margin and is input as an initial parameter to the {\sc TraP}. In Section 4.3 we address strategies that can be used to find the optimal margin for detecting transient sources.

\section{Finding variable sources via machine learning algorithms}

In the previous section, we showed that {\sc TraP} is able to identify transient and variable sources using a number of thresholds that can be fine tuned to meet the astronomer's requirements. However, in all of the cases, it is not obvious which values will give optimal results. In this section we introduce machine learning strategies that can be used to train these parameters. As discussed in the introduction, unsupervised machine learning strategies typically require  datasets containing sufficient sources in each category (and these categories are unknown). As sufficiently large datasets (containing sufficient numbers of the rare transient and variable sources) are still being produced, we choose to focus on two supervised algorithms in this analysis  with data simply labelled as `transient', `variable' or `stable'. 

There are a wide range of supervised machine learning techniques that could be applied to the data. In this paper, we have chosen to focus on two supervised machine learning techniques: simple anomaly detection (described in Sections 4.1.1 and 4.1.3) and penalised logistic regression (described in Section 4.1.2). We have chosen two relatively simple algorithms in this work as {\sc TraP} users currently focus on a small number of variability parameters for each source. This work aims to produce a good grounding for future development, when larger and more complex datasets become available.

In this Section, we describe the different algorithms used (Section 4.1), the testing strategies applied to confirm that the algorithms are obtaining their optimal solution (Section 4.2) and a discussion of identified transient and variable sources (Section 4.3).

\subsection{Algorithms used}

\subsubsection{Anomaly Detection}

We have shown that the transient parameters, $\eta_{\nu}$ and $V_{\nu}$, of stable sources follow an approximately Gaussian distribution in logarithmic space while variable sources are clearly anomalous to these distributions. Therefore, the variable and stable sources can be distinguished by using a $\sigma$ threshold cut, with all anomalous sources above the threshold being labelled transient. Although we can define a threshold on $\eta_{\nu}$ using a FDR (as typically used by transient surveys), it is clear that using this method with LOFAR data will lead to either many false detections or miss a large region of parameter space where variable sources may occur. We have suggested that this parameter space can be described using sigma threshold cuts on both the $\eta_{\nu}$ and $V_{\nu}$ values, but it is not obvious what the optimal thresholds should be. {\sc TraP} was initially designed assuming that the two simple threshold cuts on the variability parameters would adequately identify transient and variable sources, meaning that two thresholds is the easiest strategy for all {\sc TraP} users to implement without significant modelling of their data. Therefore, in this section, we describe a preliminary method, a sigma threshold, that can be applied quickly and easily to all current datasets and {\sc TraP} versions.

This problem lends itself to a simple machine learning classification algorithm such as anomaly detection \citep[first proposed by][]{Denning:1987} that divides a dataset into two categories: a large sample of ``normal'' sources and a group of ``unusual'' sources. This method requires that the ``normal'' sources can be approximated as following a Gaussian distribution, with the unusual sources being anomalous to the distribution, and we have demonstrated that our data meet this requirement. This machine learning algorithm can either be unsupervised, where an arbitrary sigma threshold is applied \citep[as utilised by ][ with 2$\sigma$ and 3$\sigma$ thresholds respectively for both parameters]{Rowlinson:2016, Stewart:2016}, or supervised (trained to give the optimal answers). Ideally, we want to choose the optimal combined thresholds for $\eta_{\nu}$ and $V_{\nu}$ so a supervised anomaly detection algorithm, trained using standard machine learning techniques, is appropriate.

Using the hybrid dataset described in Section 2.3 as our training datasets, we can determine the optimum multiples of $\sigma$ which maximise the number of real transients detected and minimise the number of spurious detections. As described in Section 3.3, a number of the simulated variable sources will be detected by the ``new source'' strategy in {\sc TraP} and do not always follow the same behaviour as the rest of the variable sources, and therefore we remove the ``new sources'' from the simulated dataset when training the thresholds. To determine the optimal thresholds, we allow $\eta_{\nu}$ and $V_{\nu}$ to have thresholds defined using different multiples of sigma. We use a range of multiples of $\sigma$ (0--3.5 $\sigma$, in 500 isotropically distributed bins to fill out the parameter space) for different thresholds on $\eta_{\nu}$ and $V_{\nu}$, giving $2.5\times10^5$ different combinations of $\sigma_{\eta_{\nu}}$ and $\sigma_{V_{\nu}}$. For each combination of thresholds, we count the number of True Positives (TP, simulated transients correctly identified), False Positives (FP, RSM sources falsely identified as transient; note that this relies on the assumption that no sources in the RSM dataset are real transients), False Negatives (FN, simulated transient sources not identified as transient) and True Negatives (TN, RSM sources correctly identified as not transient). Using these values, we calculate the precision (the probability that the transients identified are real transients, also known as reliability) and recall (the probability that all the transients in the dataset have been identified, commonly also referred to as completeness) using 

\begin{eqnarray}
{\rm Precision} = \frac{{\rm TP}}{{\rm TP}+{\rm FP}} \equiv 1 - FDR \label{eqn:precision}, \\
{\rm Recall} = \frac{{\rm TP}}{{\rm TP}+{\rm FN}} \label{eqn:recall}.
\end{eqnarray}

For reference, we calculate the precision and recall obtained for the hybrid dataset using thresholds from \cite{Rowlinson:2016} and \cite{Stewart:2016}. \cite{Rowlinson:2016}, with thresholds of $2\sigma$ on both $\eta_{\nu}$ and $V_{\nu}$, attains a precision of 54\% and a recall of 80\%. While \cite{Stewart:2016}, with thresholds of $3\sigma$ on both $\eta_{\nu}$ and $V_{\nu}$,  attains a precision of 95\% and a recall of 26\%.

\begin{figure}
\centering
\includegraphics[width=0.48\textwidth]{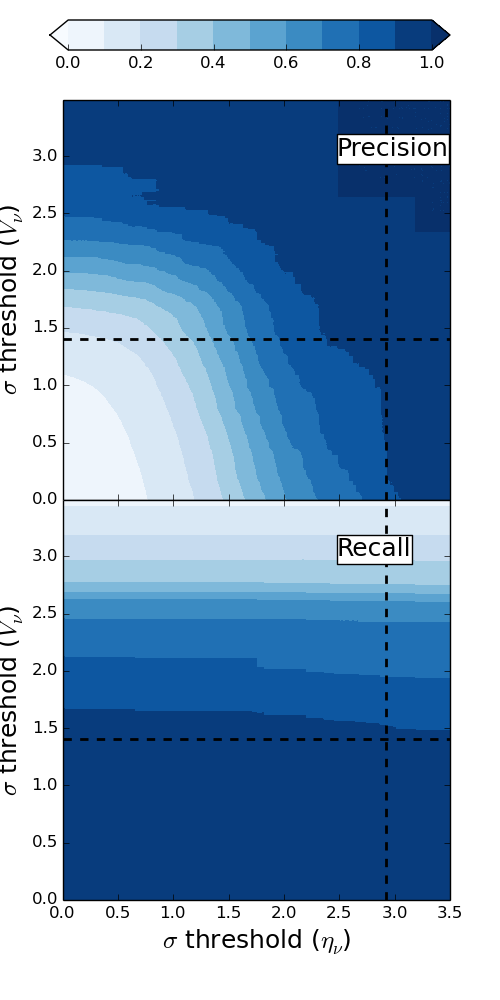}
\caption{The precision and recall (as defined in Equations \ref{eqn:precision} and \ref{eqn:recall}) as a function of the $\sigma$ threshold used. The dashed lines show the trained thresholds (as described in the text), attaining a precision of 95\% and recall of 93\%.}
\label{fig:sim_precisions}
\end{figure}

In Figure \ref{fig:sim_precisions}, we show the precision and recall for the simulated variables as a function of the two $\sigma$ thresholds used to identify anomalous sources. The contour colour scale shows the probabilities with the darkest colours giving the highest probabilities that the sources are reliably identified as variable (precision) or that all the variable sources are recovered (recall). The contours are obtained by conducting a cubic interpolation across a grid between the data points \citep[using the interpolate function in the {\sc Python} module {\sc SciPy};][]{Jones:2001} and these contours can be used to find the optimal $\sigma$ thresholds. By inputting required values for the precision (consistent with a FDR) and recall, the contours can be used to determine the $\sigma$ thresholds that should be used. The training algorithm finds the $\sigma$ thresholds that can attain the required precision and recall using the following method:
\begin{enumerate}
    \item select all combinations in the grid of trialled thresholds that exceed the required precision.
    \item calculate the 2D linear separation of the input requirements and the combinations remaining from the first step.
    \item identify the combination of sigma thresholds that gives the closest match to the input precision and recall.
    \item if an exact match cannot be obtained, the algorithm finds the best match to the input precision and the highest recall value this corresponds to.
\end{enumerate}

To demonstrate this technique we choose to apply a precision and a recall of 95\% to the combined observed and simulated dataset. The algorithm determines that the thresholds should be as follows: $\sigma_{\eta, {\rm threshold}} = 2.93$ and $\sigma_{V, {\rm threshold}} = 1.40$. Note that these parameters are dependent upon the range of SNRs used for the simulated sources. Therefore, to attain the best thresholds for a dataset, it is important to retrain using simulated sources closest to those expected in the dataset (for an unknown population, use a range of values as used in this analysis). These $\sigma$ thresholds, illustrated with the dashed lines in Figure \ref{fig:sim_precisions}, give a precision of 95\% and recall of 93\% for the training data, almost meeting the inserted requirements for precision. The method does not quite reach the required recall of 95\% as it is not possible to attain both a high precision and recall with this dataset; however the method chooses the closest match to the input requirements, prioritising by precision when required. 

\begin{figure}
\centering
\includegraphics[width=0.5\textwidth]{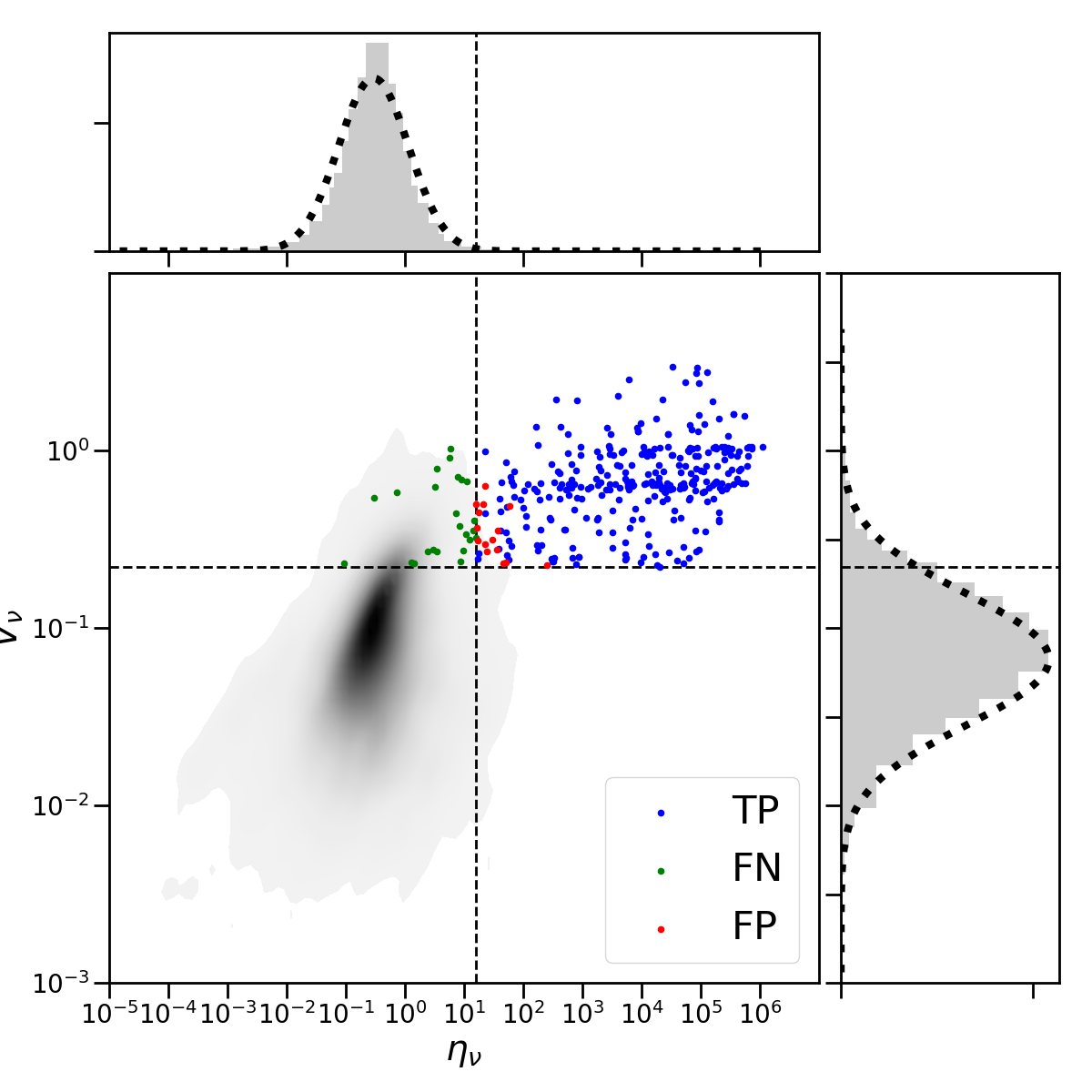}
\caption{The results of classifying the simulated and RSM datasets, using the threshold as given in Figure \ref{fig:sim_precisions}, plotted by their transient parameters. The colour scheme, as given in the legend, shows the True Positives (correctly identified transients), False Positives (stable sources incorrectly identified as transient), and False Negatives (transients not found). The True Negatives (stable sources correctly identified) are shown as the shaded region. The dashed lines show the $\eta_{\nu}$ and $V_{\nu}$ thresholds attained using the $\sigma$ thresholds illustrated in Figure \ref{fig:sim_precisions}.}
\label{fig:sim_results2}
\end{figure}

\begin{figure}
\centering
\includegraphics[width=0.5\textwidth]{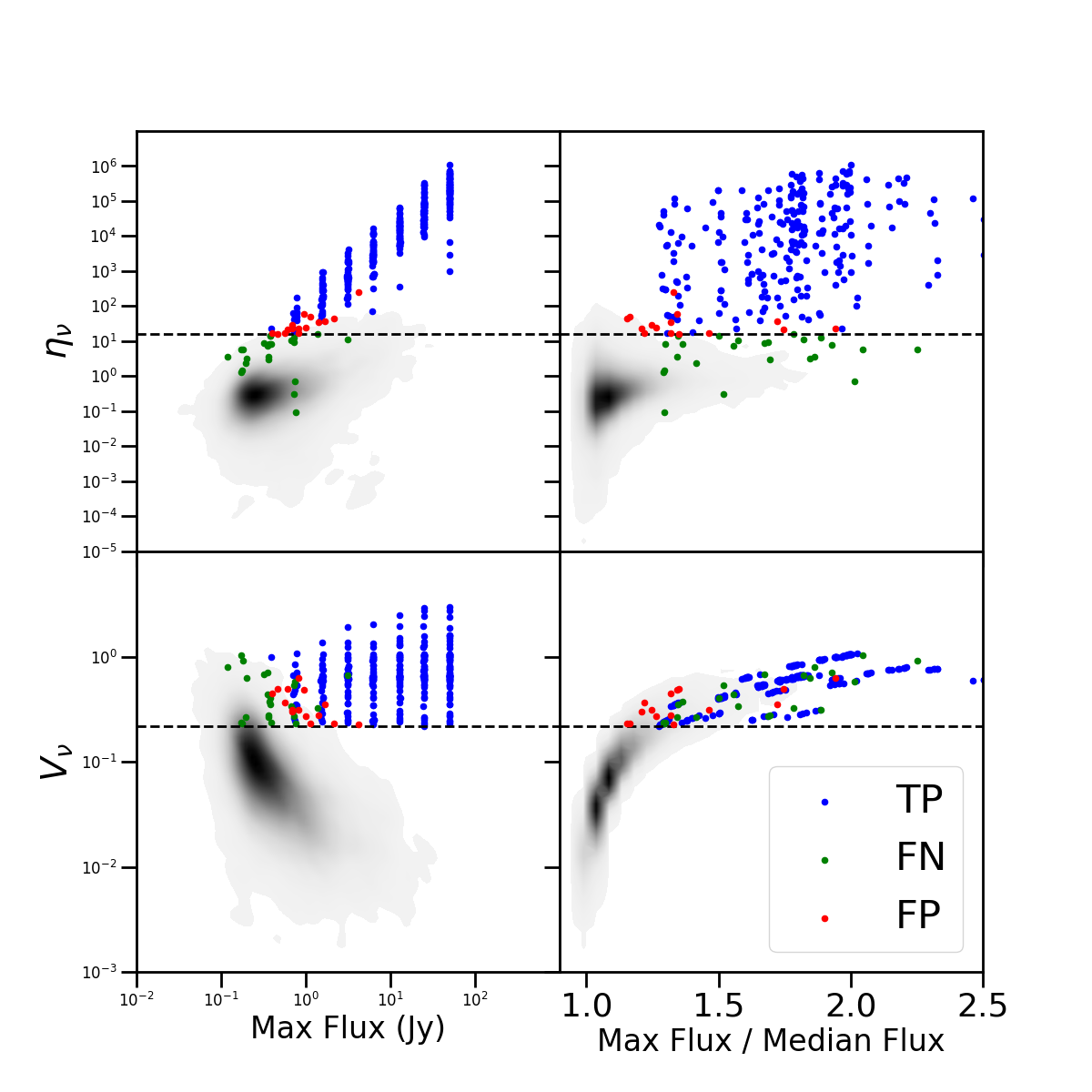}
\caption{As in Figure \ref{fig:rsm_trans_params2}, the transient parameters are plotted for all the sources observed in the simulated datasets (colour scheme in legend and as in Figure \ref{fig:sim_results2}). The dashed lines show the $\eta_{\nu}$ and $V_{\nu}$ thresholds attained using the $\sigma$ thresholds illustrated in Figure \ref{fig:sim_precisions}.}
\label{fig:sim_results2b}
\end{figure}

In Figure \ref{fig:sim_results2} and \ref{fig:sim_results2b}, we show the identifications resulting from classification using a precision of 95\% and a recall of 95\%. As noted previously, it was not possible to precisely attain these values for this dataset and the algorithm has determined a `best match' with a precision of 95\% and a recall of 93\%. A number of transient sources are not identified (FN, false negatives) and, by cross-matching with Figure \ref{fig:sim_trans_params2}, these are typically the sources that have a maximum flux density ratio $\lesssim$2 (i.e. the sources which do not have a large amplitude variability). The majority of transient sources with peak flux densities $\gtrsim$1 Jy are detected and below this limit they are indistinguishable from the stable sources so we are unlikely to identify these sources. We note that this is a significantly better performance than the arbitrary thresholds chosen by \cite{Rowlinson:2016} and \cite{Stewart:2016}.

\subsubsection{Logistic Regression}

In Section 4.1.1, we showed how to determine detection thresholds using a sigma threshold determined from the roughly Gaussian distribution of $\eta_\nu$ and $V_\nu$ for stable sources. This anomaly detection strategy, a very simple machine learning strategy, can be tuned to minimise the number of false detections; but it could be improved by using a single diagonal line that is a function of $\eta_{\nu}$ and $V_{\nu}$, instead of the horizontal and vertical thresholds. Furthermore, this anomaly detection strategy does not utilise all the available parameters for each source. For instance, in Figure \ref{fig:sim_trans_params2}, the transient sources are clearly separated from the stable sources when also considering the maximum flux density and maximum flux density ratio of the sources. Although it is possible to extend this anomaly detection strategy to include more parameters, a simple threshold strategy would not be appropriate for all of the available parameters (e.g. the maximum flux). Additionally, the training would become increasingly complex resulting in a further significant slow-down in training time. In this section, we utilise a logistic regression algorithm \citep[a supervised machine learning strategy; e.g.][]{Darroch:1972} to classify datasets, initially trained using categorised data and then applied to new data points to determine which category they belong in\footnote{We apply training obtained in this method from the Stanford Machine Learning course available on Coursera (https://www.coursera.org/course/ml)}. In the simplest case, with two features and two categories, this strategy separates the two datasets using a straight line. Here, we use the same labeled dataset as in Section 4.1 and four features; $\eta_\nu$, $V_\nu$, maximum flux density ($F_{\rm max}$) and maximum flux density ratio: $R=\left( \frac{\rm Max~Flux}{\rm Average~Flux} \right)$. These features are given in the matrix equation\footnote{where matrices are denoted using bold font symbols.} 
\begin{eqnarray}
\mathbf{X} = \begin{bmatrix}
 1 & \log_{10}(\eta_\nu) & \log_{10}(V_\nu) & \log_{10} (F_{\rm max}) & R\\
 \vdots & \vdots & \vdots & \vdots & \vdots 
 \end{bmatrix}, \label{eqn:X}
\end{eqnarray}
where each source is identified as `stable' (0) or `variable' (1) by a corresponding element in the matrix
\begin{eqnarray}
\mathbf{y} = \begin{bmatrix}
 1 ({\rm or~0}) \\
 \vdots
 \end{bmatrix}. \label{eqn:y}
\end{eqnarray}
The logistic regression method deployed then separates the `stable' and `transient' sources using a linear algorithm in 4-dimensional space, $\mathbf{\theta} \bullet \mathbf{X}$, where the model parameters (one for each of the features in Equation \ref{eqn:X}) are
\begin{eqnarray}
\boldsymbol{\theta}=\begin{bmatrix}
\theta_0 \\
\theta_1 \\
\theta_2 \\
\theta_3 \\
\theta_4
\end{bmatrix} \label{eqn:theta}
\end{eqnarray}
(with 1 parameter, this would simplify to a straight line defined by $\theta_0 + \theta_1 \eta_\nu$). The model predicts the classification of a source by calculating
\begin{eqnarray}
\boldsymbol{\Sigma} = (1+e^{-\boldsymbol{\theta} \bullet \mathbf{X}})^{-1},
\label{eqn:logistic}
\end{eqnarray}
which outputs $\sim$1 if the source is predicted to be `variable' or $\sim$0 if the source is predicted to be `stable'. The model parameters then require training, which is conducted by minimising the difference between the labels (given in Equation \ref{eqn:y}) and the predictions from Equation \ref{eqn:logistic}, using
\begin{eqnarray}
\mathbf{J} = \frac{1}{N} \left(-\mathbf{y} \bullet \log_{10}(\boldsymbol{\Sigma}) - (1-\mathbf{y}) \bullet \log_{10}(1-\boldsymbol{\Sigma}) \right) \nonumber \\
+ \frac{\lambda}{2N}\sum\limits_{i=0}^4  \boldsymbol{\theta}_i^2, \label{eqn:J}
\end{eqnarray}
where $N$ is the number of sources processed, $\lambda$ is the regularisation term which can be used to bias the solution against over-fitting the data with multiple features and $\bullet$ represents the matrix dot product. We find the best classification by minimising Equation \ref{eqn:J} using the optimise function in the {\sc Python} module {\sc SciPy} \citep{Jones:2001}. 

\begin{figure}
\centering
\includegraphics[width=0.5\textwidth]{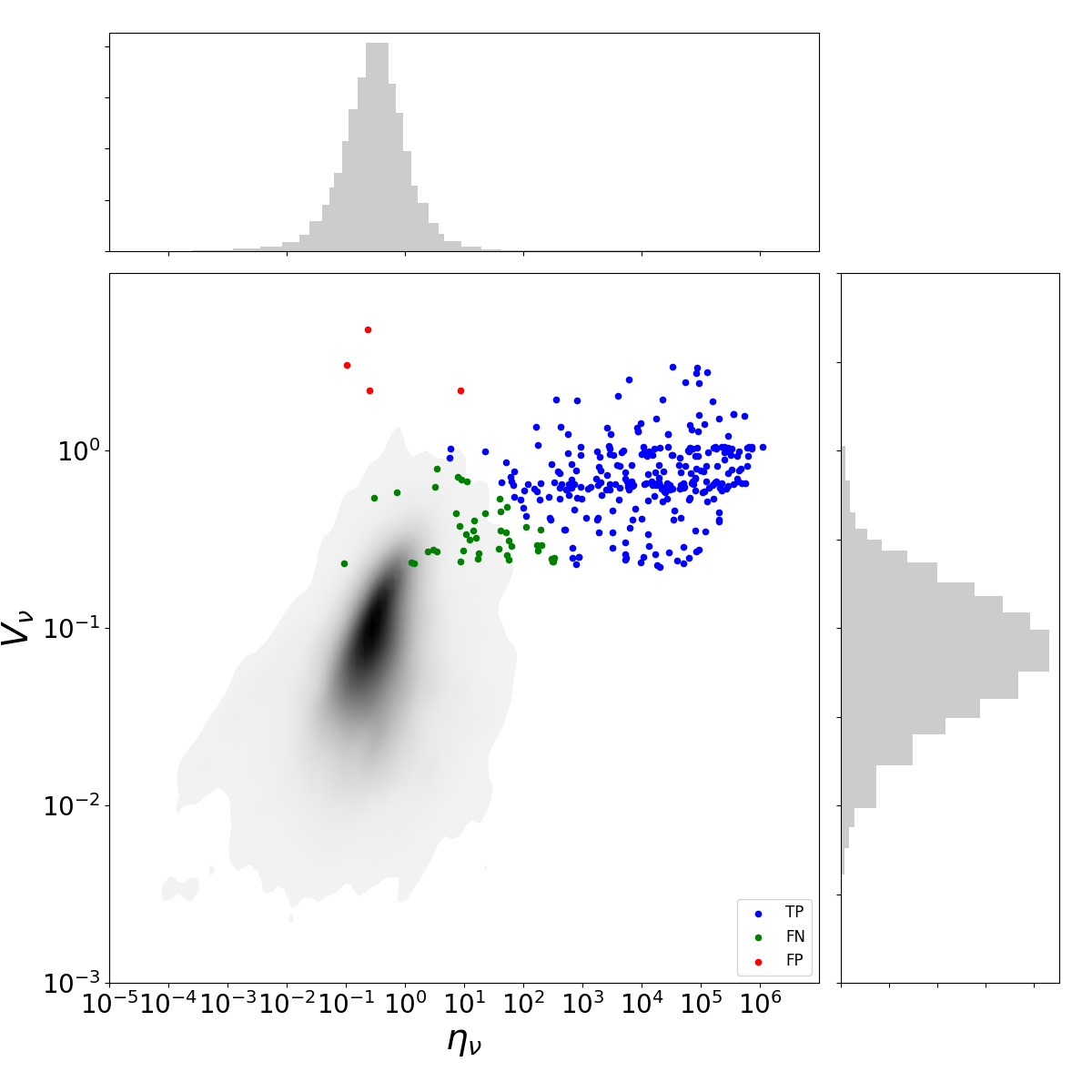}
\caption{The transient parameters, $\eta_{\nu}$ and $V_{\nu}$, for each of the data points classified using the trained logistic regression algorithm (color scheme as in Figure \ref{fig:sim_results2}).}
\label{fig:machine_learning_results1}
\end{figure}

\begin{figure}
\centering
\includegraphics[width=0.5\textwidth]{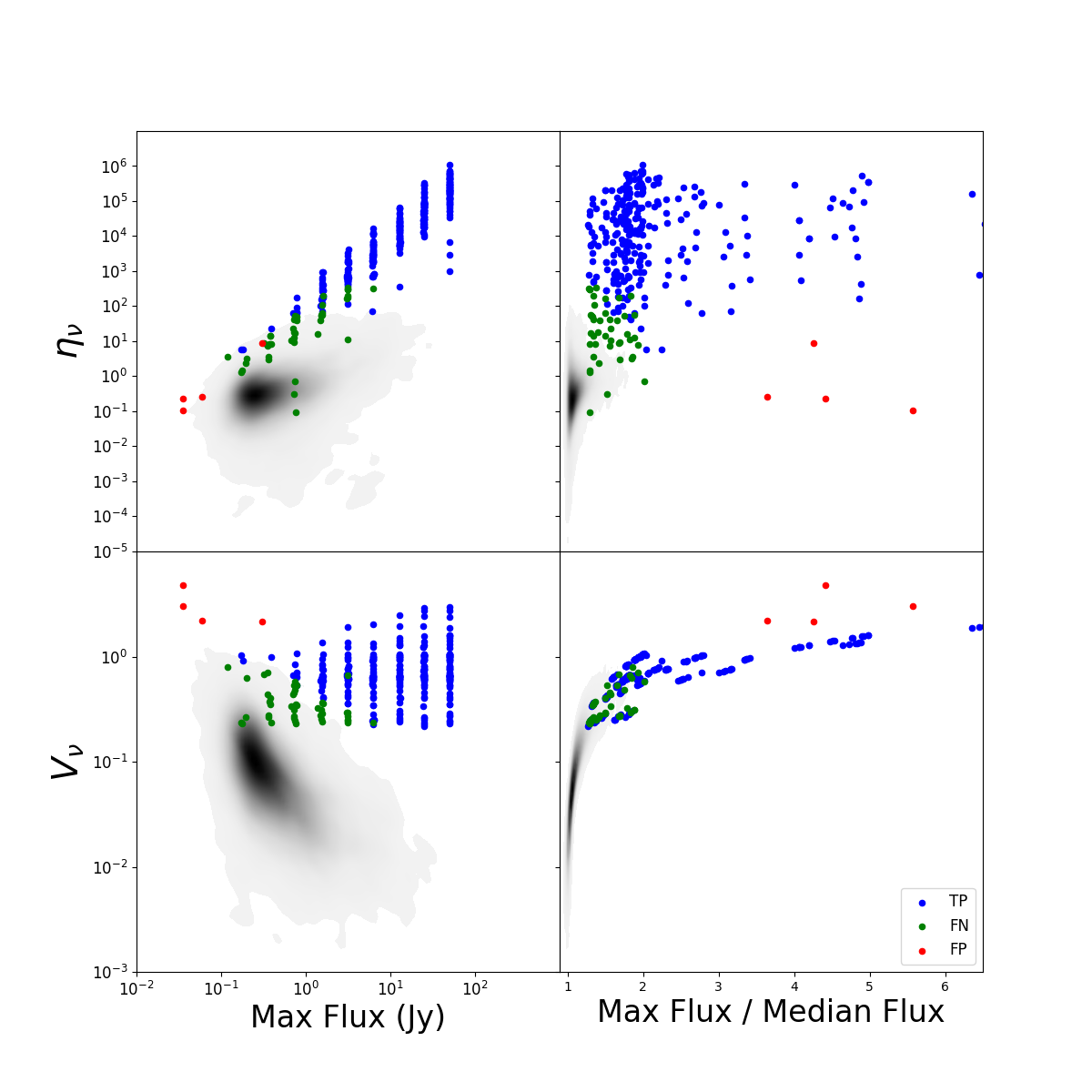}
\caption{All the features used to train the algorithm with the data points classified using the trained logistic regression algorithm (color scheme as in Figure \ref{fig:sim_results2}, otherwise as Figure \ref{fig:rsm_trans_params2})}
\label{fig:machine_learning_results2}
\end{figure}

Therefore, we train the algorithm on the full dataset to obtain the optimum parameters and then classify all the available data to determine the precision and recall of the algorithm (as defined in Equations \ref{eqn:precision} and \ref{eqn:recall}, n.b. see Section 4.2.2 for the testing of this algorithm). We find the precision is 98\% with a recall of 91\% for the full dataset, providing an excellent identification of transient sources. In Figures \ref{fig:machine_learning_results1} and \ref{fig:machine_learning_results2}, we show the classification results using this logistic regression algorithm. We note that a number of sources give false negative results, these are typically sources that are not significantly variable in comparison to the stable sources.

After the initially time consuming training step, classification takes $<15$ s for $\sim$6700 sources making it suitable for real-time analysis. It is technically very simple to introduce new features in the logistic regression algorithm and further work is required to identify new features from the light curves, and potentially multi-wavelength catalogue data where available, which can significantly improve the classification. Additionally, to confirm the performance of the method and obtain better constraints on the precision and recall, this algorithm needs testing on much larger datasets which are drawn from different samples other than the RSM.

\subsubsection{Training transient source identification}

As stated in Section 3.4, the transient sources are identified in comparison to the detection threshold in the previous best image plus a margin. In this section, we demonstrate how the optimal margin can be identified using the simulated transients and the candidate transients identified within the RSM dataset. As before, we assume that there are no real transient sources within the RSM dataset.

\begin{figure}
\centering
\includegraphics[width=.49\textwidth]{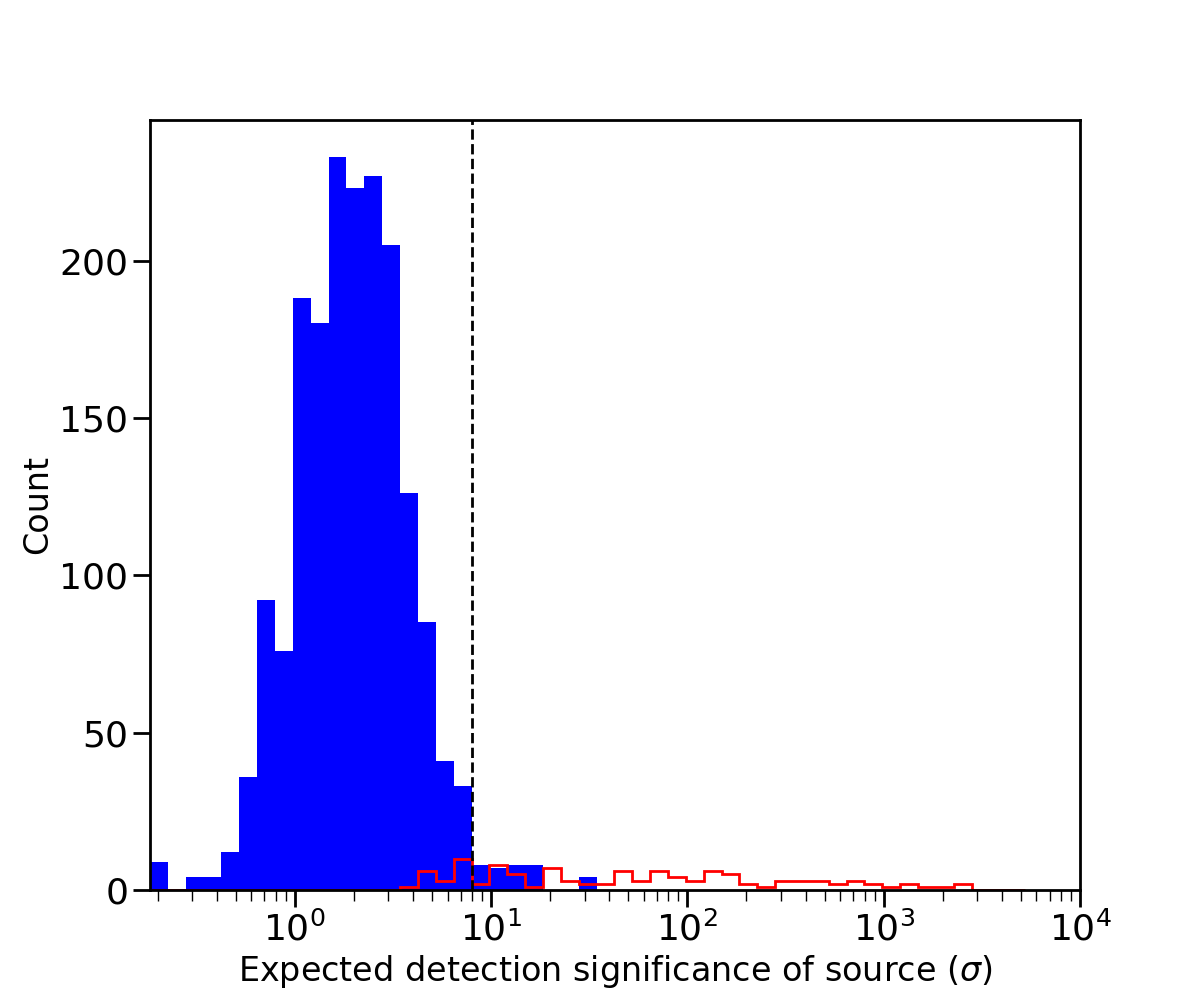}
\includegraphics[width=.49\textwidth]{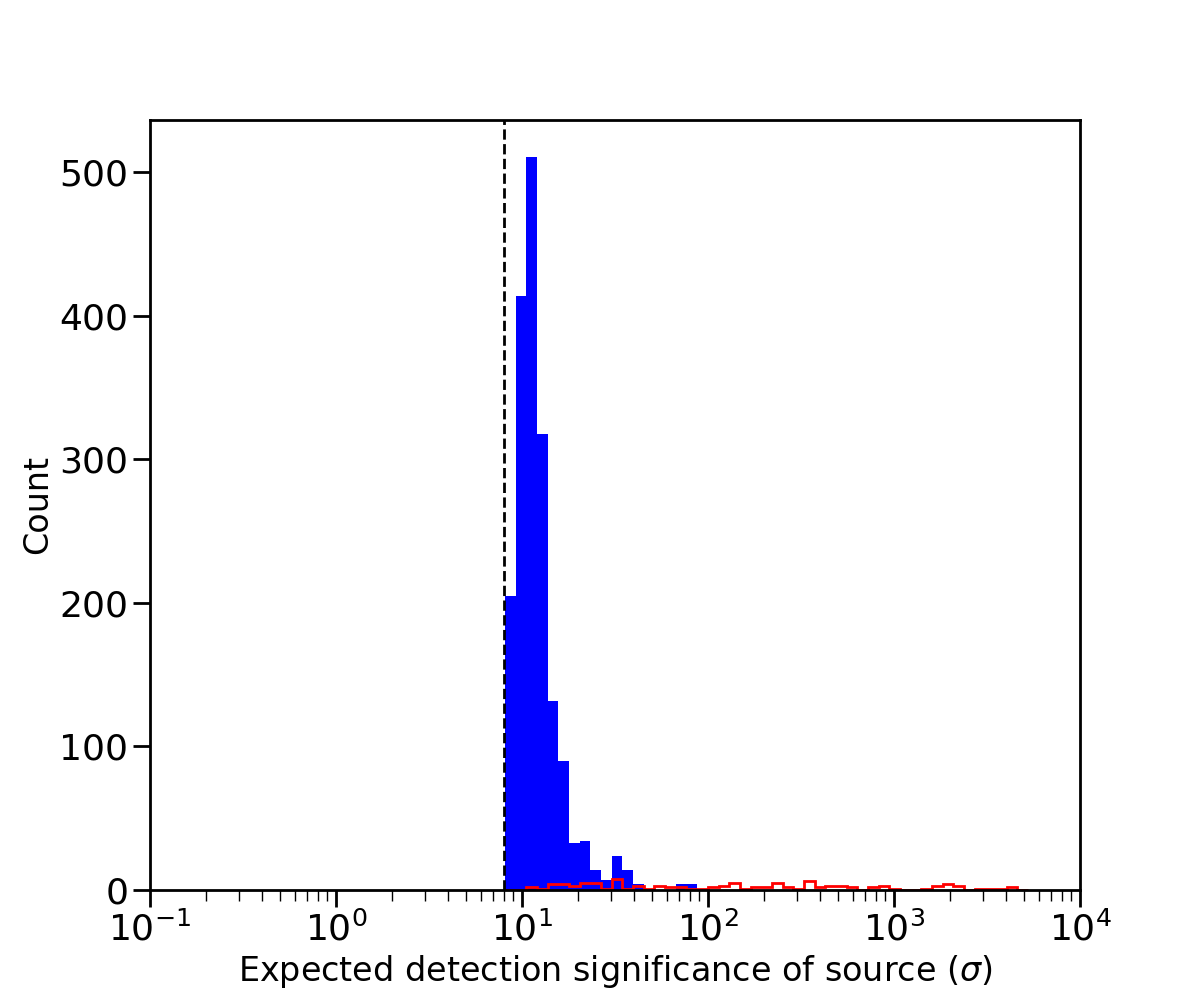}
\caption{Histograms showing the number of sources with an expected detection significance $\sigma$ from the best (top) and worst (bottom) rms regions from the best previously analysed image. The dashed black line shows the detection threshold of 8 $\sigma$ used to extract sources. In blue are the newly detected sources from the RSM and overlaid in red are the simulated transient sources.}
\label{fig:rms}
\end{figure}

For each newly identified source, {\sc TraP} determines the signal to noise that the source would have in the best and worst parts of the previously best image of the field ($\sigma_{{\rm rms,best}}$ and $\sigma_{{\rm rms,worst}}$ respectively). These values are then compared to the detection threshold utilised, 8$\sigma$ in this analysis, plus a margin and two outcomes are determined:
\begin{itemize}
  \item $\sigma_{{\rm rms,worst}} > ({\rm detection~threshold}~+~{\rm margin}) $: the source is a transient
  \item $\sigma_{{\rm rms,best}} > ({\rm detection~threshold}~+~{\rm margin}) $: the source is a transient candidate
\end{itemize}
In Figure \ref{fig:rms}, we show the distributions of the $\sigma_{{\rm rms,best}}$ and $\sigma_{{\rm rms,worst}}$ for the new sources in the RSM dataset and the simulated transient sources.

To determine the optimal margin to be added to the detection threshold, we trial multiple margins in the range 0--50$\sigma$. For each margin, we calculate the precision and recall using Equations \ref{eqn:precision} and \ref{eqn:recall} assuming that the new sources from the RSM dataset are not transient. The optimal margin is then defined as that which gives the required precision and recall. In this scenario, we chose the optimal margin to be that which maximises the ``F-Score'' given by

\begin{eqnarray}
{\rm F\mbox{-}Score} = \frac{2 \times {\rm precision} \times {\rm recall}}{{\rm precision}+{\rm recall}} \label{eqn:Fscore}.
\end{eqnarray}

\begin{figure}
\centering
\includegraphics[width=.49\textwidth]{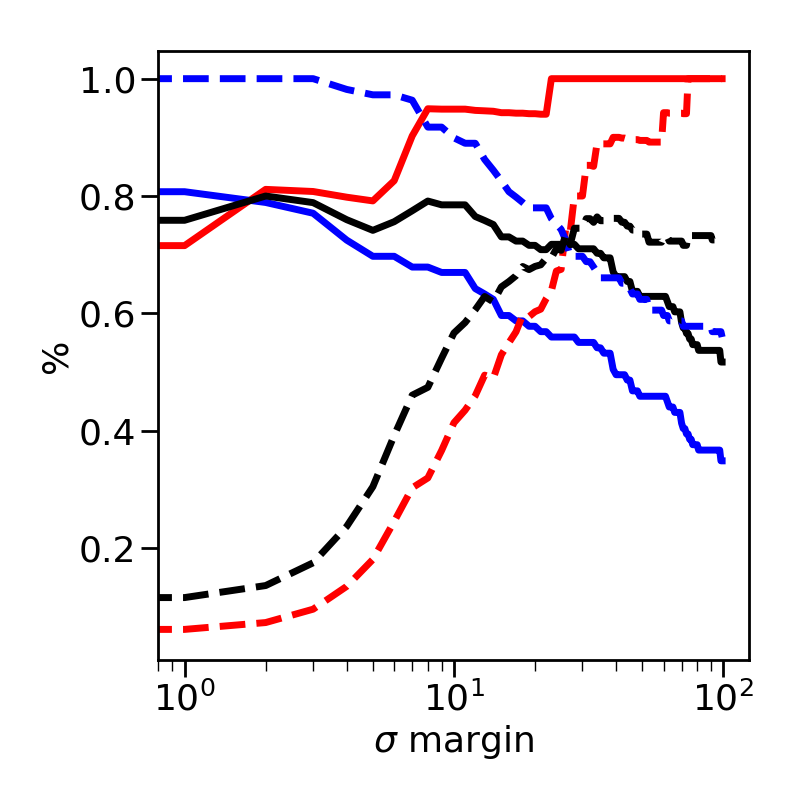}
\caption{This figure shows how the precision (red), recall (blue) and F-Score (black) for the transient source classification change as the margin added to the detection threshold is increased. The solid lines are the results for the worst rms region in the previous best image and the dashed lines are for the best rms region. We chose the best parameter to maximise the F-Score giving an optimal balance between precision and recall.}
\label{fig:bestmargin}
\end{figure}

In Figure \ref{fig:bestmargin} we plot the precision, recall and F-score for the best and worst rms regions. When using the worst rms from the best image, we find that the optimal margin is 2$\sigma$. In this scenario, the precision is good at 81\% with a recall of 79\%. Alternatively, when using the best rms from the best image we find that a margin of 34$\sigma$ gives a higher precision of 89\% and lower recall of 67\%. Therefore, it is up to the astronomer to determine which scenario meets their requirements.

\subsection{Testing of algorithms}

Each of the machine learning algorithms, are tested using the methods described in this section. Some of the tests are specific to each of the methods (Sections 4.2.1 -- 4.2.2), but we also include comparative tests where possible to compare the different strategies in Sections 4.2.3 -- 4.2.4.

When testing the algorithms we subdivide the dataset into smaller samples. A randomised 60\% of the available dataset was used to train the algorithm, 30\% was used for cross validation to ensure a good solution and the remaining 10\% was used as a test dataset which was used to manually confirm that the classification was working successfully. To compare the results obtained, we use the training error (the error in classifying the training dataset, blue in Figures \ref{fig:validation} -- \ref{fig:trials}) and the validation error (the error in classifying the validation dataset, green in the Figures), where the error is defined as:
\begin{eqnarray}
{\rm error} =  \frac{\sum|{\rm input~category} - {\rm predicted~category}|}{\rm total~number~of~sources} \nonumber \\
\equiv \frac{{\rm FP} + {\rm FN}}{{\rm FP} + {\rm FN} + {\rm TP} + {\rm TN}} \label{eqn:err}
\end{eqnarray}
and can also be referred to as the fraction of incorrect classifications.

\subsubsection{Anomaly Detection specific tests}

The anomaly detection strategy has been adapted from an unsupervised technique to a supervised one by measuring the precision and recall for multiple combinations of the sigma thresholds with the observed data. The resulting dataset is then used to find the closest value to the input requirements. To confirm that the anomaly detection strategy identifies the thresholds correctly, we trial a range of input precision and recall values (in the range 0.7--0.95). The output precision values are equal to the requirements input confirming this strategy is working effectively. The recall values are slightly more variable, but lie within a few percent of the input requirements.

\subsubsection{Logistic Regression specific tests}

\begin{figure}
\centering
\includegraphics[width=0.4\textwidth]{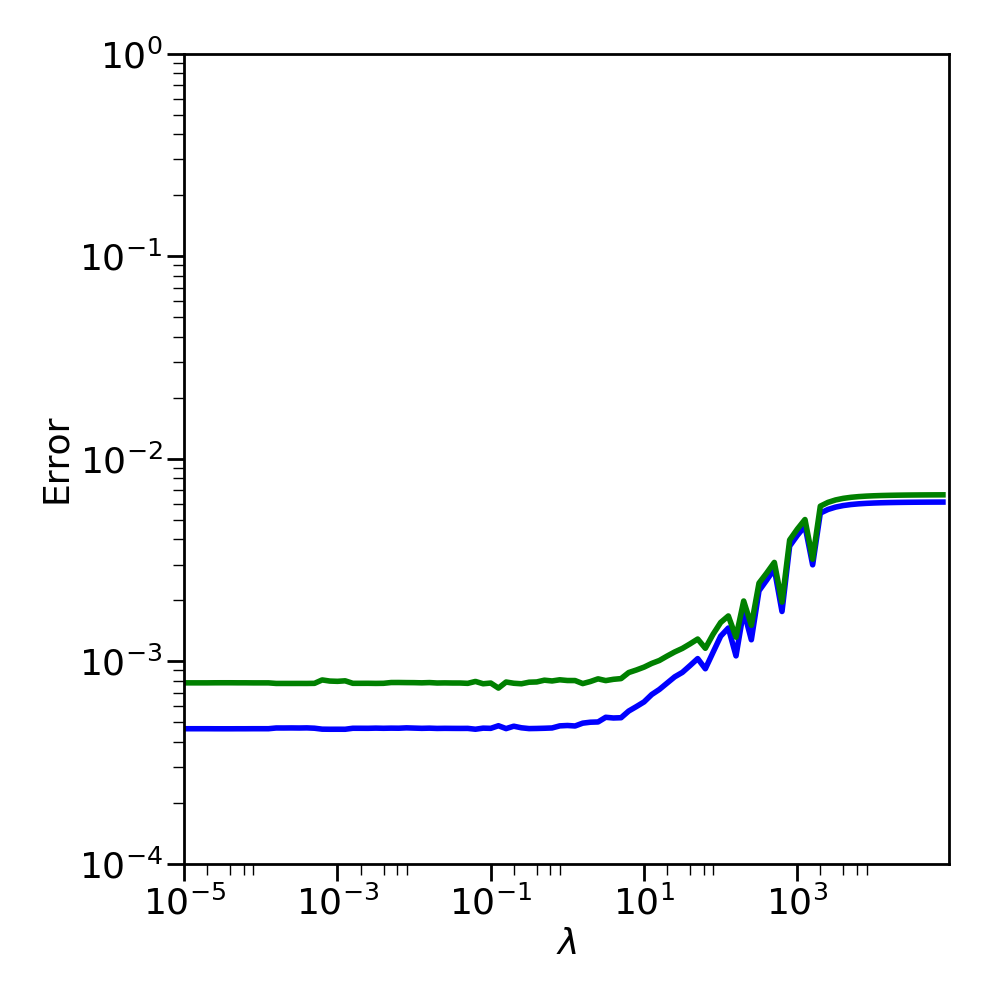}
\caption{The effect of varying the $\lambda$ regularisation parameter used on the training (blue) and validation (green) errors: high $\lambda$ values suppress some of the model parameters, $\theta$, which can prevent overfitting of the data. As we have sufficient data and tend to a solution we can set a low value for $\lambda$.}
\label{fig:validation}
\end{figure}

In the logistic regression algorithm, there is a tune-able parameter $\lambda$ which can help prevent over-fitting the data with a model that has too many free parameters for the observed dataset. We tested our choice of $\lambda=0.1$ in the machine learning algorithm by retraining the dataset using a range of $\lambda$ values ($10^{-5}$ -- $10^{5}$). We show the training and validation error for the different $\lambda$ values in Figure \ref{fig:validation}. For large values of $\lambda$, the training and validation errors are relatively high, but for $\lambda<10$ we find that the errors do not change significantly and they fall to below $10^{-3}$ suggesting the classification is very accurate. Therefore, our choice of $\lambda=0.1$ is appropriate for this dataset and we note that a range of $\lambda$ values could be used to successfully train this dataset. Additionally, we trialled $\lambda = 0$ and it produced a near identical solution to $\lambda = 0.1$. Therefore, for this specific dataset, the $\lambda$ term is not strictly required. However, as this method is expected to be extended to larger datasets containing more features, the $\lambda$ parameter will remain necessary.

\subsubsection{Learning curves}

\begin{figure}
\centering
\includegraphics[width=0.4\textwidth]{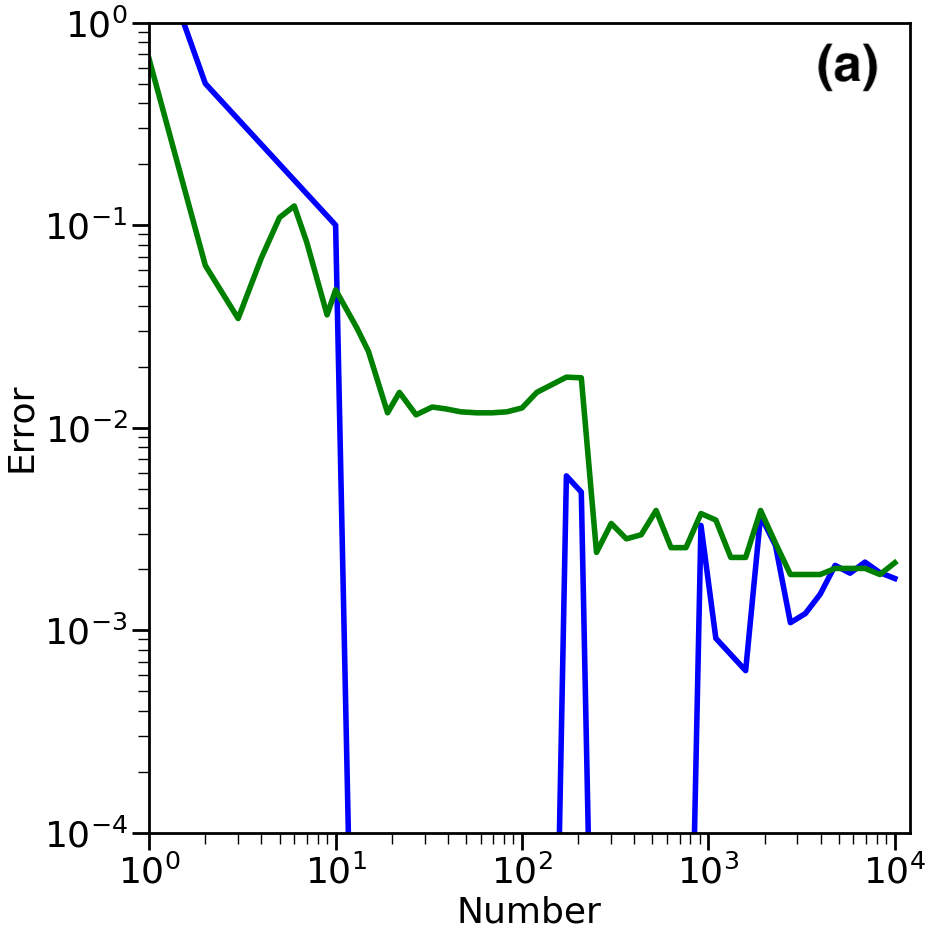}
\includegraphics[width=0.4\textwidth]{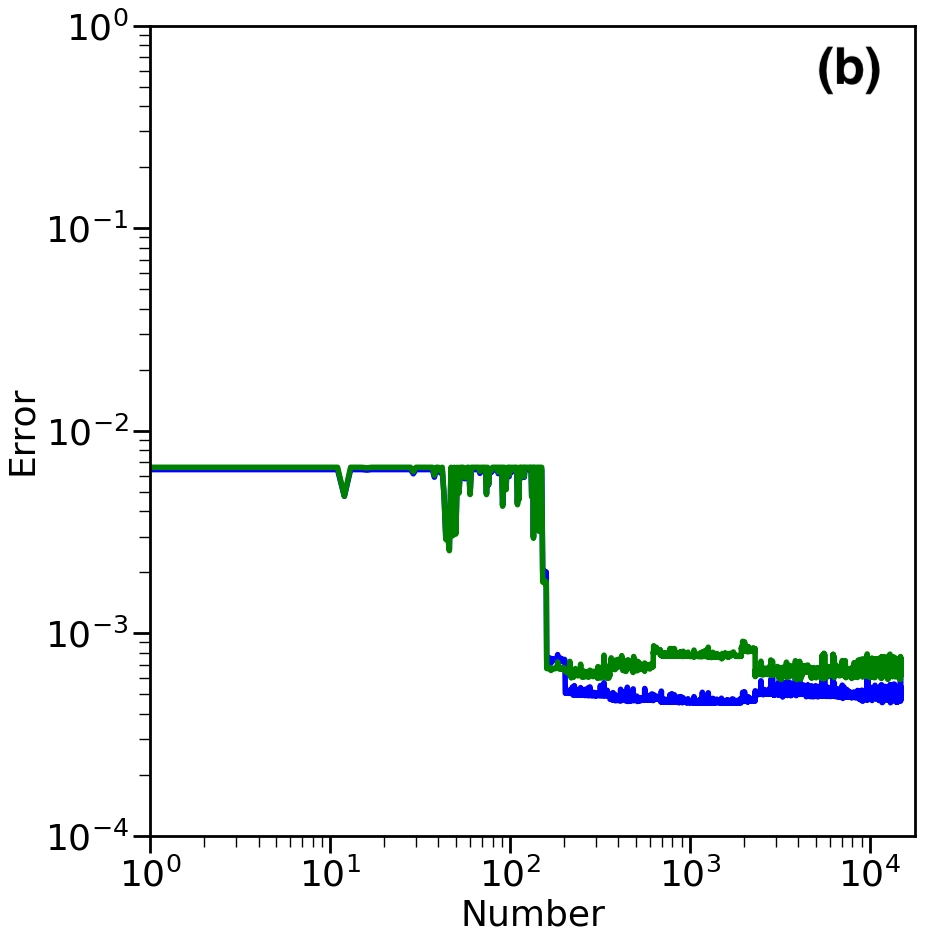}
\includegraphics[width=0.4\textwidth]{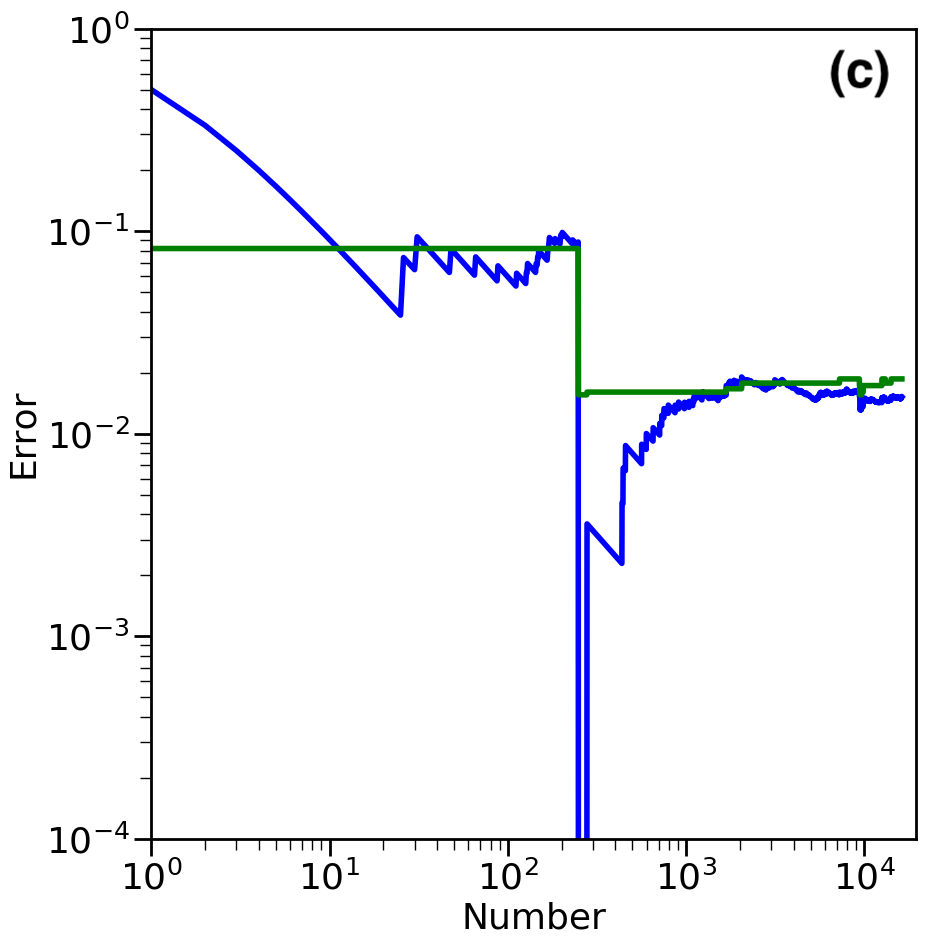}
\caption{These panels show the learning curve, the training (blue) and validation (green) errors as a function of the number of data points in the training dataset. An optimum solution is reached when the values have tended towards similar values. (a) a simplified version for the anomaly detection strategy, (b) the logistic regression algorithm using $\lambda = 0.1$, (c) the transients algorithm}
\label{fig:learning}
\end{figure}

To confirm that the algorithms are correctly tending towards a good solution, we plot the learning curve. To produce the learning curve the algorithms are systematically trained on a dataset starting with 1 data point, then increasing the dataset by 1 data point and retraining until all the data points are used. Each solution is validated by classifying the cross validation dataset and calculating the validation error. The optimum solution is obtained when the training and cross validation curves have tended to a similar value.

We directly apply this testing method to the logistic regression and the transient source identification strategy as outlined in Section 4.1.3. However, the anomaly detection strategy is extremely slow, typically taking hours to produce the gridded sigma values required (including significant parallelisation within the code). This duration could be reduced by decreasing the number of trials in the sigma grids, but this reduction in resolution has a significantly detrimental affect on the classification capability of the algorithm and would not produce representative results. Therefore, to produce a full learning curve for this strategy would take a disproportionate amount of time. To make this a more achievable option and representative of the final result, we instead trial 50 combinations logarithmically distributed in the range $1$--$1\times10^4$ data points.

The anomaly detection learning curve, in Figure \ref{fig:learning}(a), shows that the algorithm is tending towards an optimal solution after approximately 1000 data points, but there is still variation in the errors. This pattern is also observed in Figure \ref{fig:learning}(c) for the transient identification strategy. The learning curve for the logistic regression strategy, in Figure \ref{fig:learning}(b), clearly shows that an optimum solution is reached when there are $>$200 data points. However, this also shows that the solution obtained by the logistic regression algorithm used will not be further improved by adding an order of magnitude more data points to the training set. We have partially corrected for this by using a low value for $\lambda$ ($0.1$, obtained in Section 4.2.2), but  this is likely caused by the slight overlap between transients and stable sources in the parameter space given by the four features considered. Ideally we need to identify more features that can be used for transient search to improve the classification.

\subsubsection{Validation curves}

\begin{figure}
\centering
\includegraphics[width=0.4\textwidth]{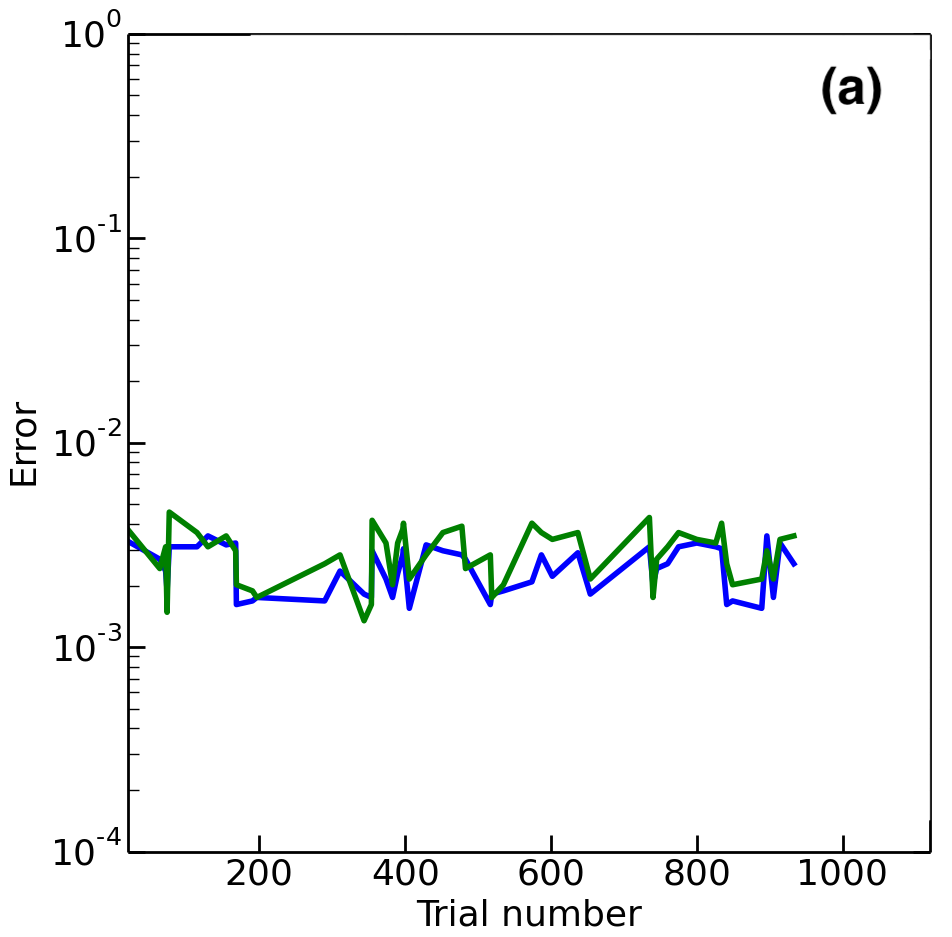}
\includegraphics[width=0.4\textwidth]{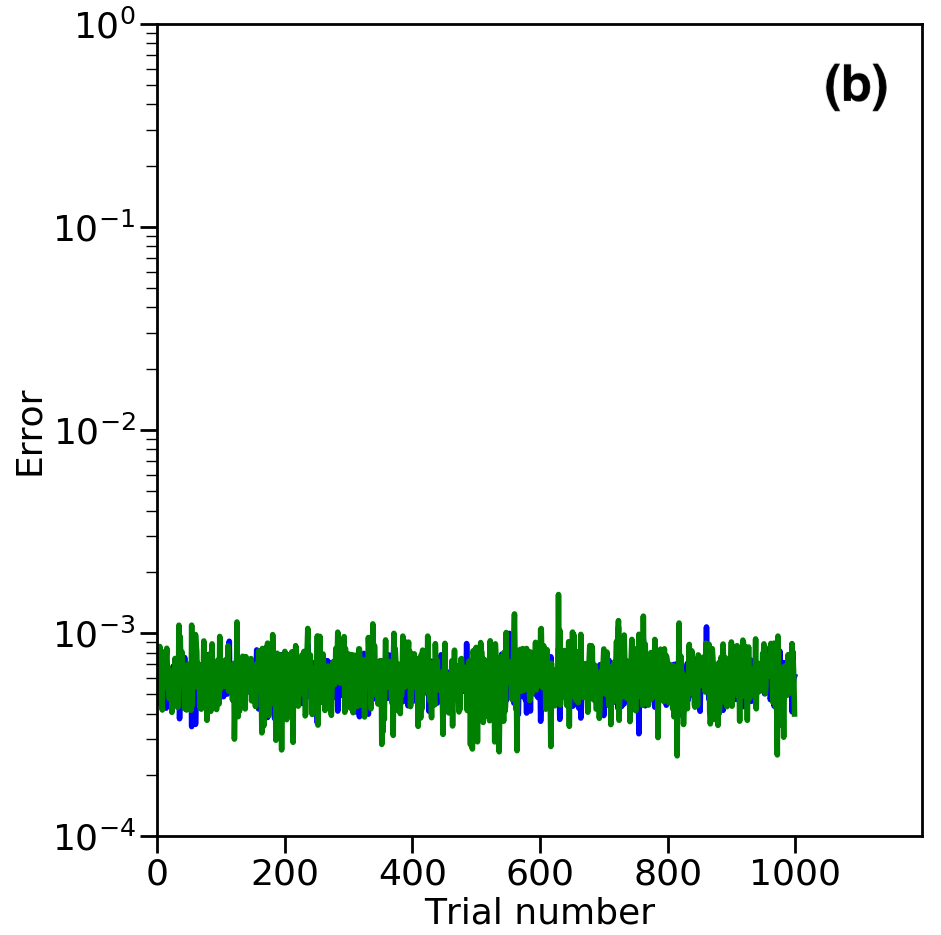}
\includegraphics[width=0.4\textwidth]{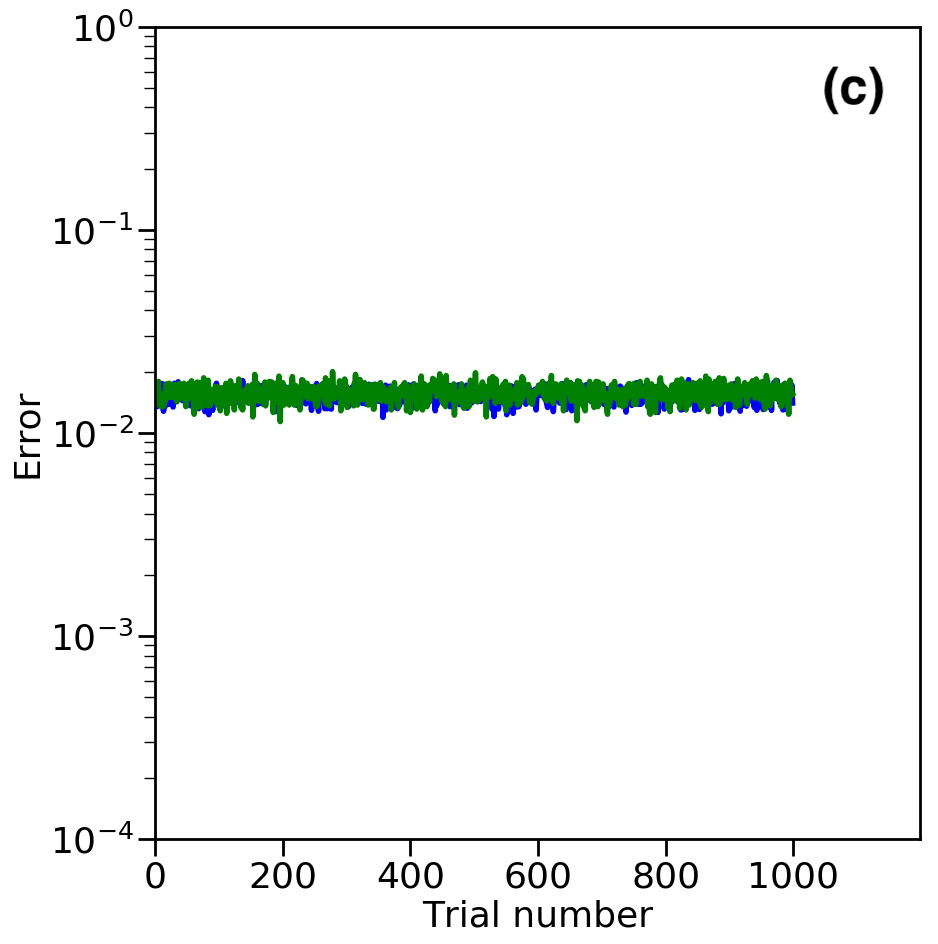}
\caption{The effect of changing the training data set on the training (blue) and validation (green) errors: we repeat the training on 1000 different random samples of 50\% of the full dataset. As the errors do not vary significantly, we can conclude that the solution is not dependent on a specific training set. (a) a simplified version for the anomaly detection strategy, (b) the logistic regression algorithm, (c) the transients algorithm}
\label{fig:trials}
\end{figure}

Additionally, we want to ensure that the solutions obtained are not caused by randomly choosing an unusual sample of the dataset to train and test the algorithm. To check this, we repeated the randomised sampling of the dataset (into the training, validation and testing datasets) 1000 times and retrained the different machine learning algorithms for each combination. As with the learning curve in Section 4.2.3, the anomaly detection strategy is too slow for this test to be completed in a practical time scale. Therefore, we randomly sample 50 of the 1000 unique datasets as a representative sample Note, in production mode for comparable datasets, these tests do not require repetition or a small subsample can be used to confirm these results). In Figure \ref{fig:trials}, we show the training and validation errors obtained for each of the 1000 different training sets for each of the machine learning algorithms. As the errors do not significantly vary for all of the different datasets and each of the training algorithms, we can conclude that the algorithm results are not dependent on the random dataset chosen to train them. We also note that the errors are significantly lower for the logistic regression strategy, showing it is outperforming the anomaly detection strategy for identifying variable sources. The errors are much higher for the transient detection strategy, this is unsurprising due to the significant overlap between simulated transients and the newly detected sources in the observed dataset (see Figure \ref{fig:rms}).

\subsection{Transients and Variables identified}

\subsubsection{Anomaly Detection Variable Candidates}

Using the trained anomaly detection strategy, requiring a precision and recall of 95\%, we identified 14 false positive sources in the RSM dataset. These false positives are variable source candidates warranting further investigation. Nine of these sources originated from a single pointing in the first run of the RSM (where each pointing of the RSM consists of 6 beams) and three candidates are from a single target field in the dataset. These twelve variable candidates were clearly caused by errors in the position or flux density scale originating from a poor calibration which had been missed by the automated image quality control settings as their image properties, such as the rms noise, were consistent with good images. Variability searches are good at identifying calibration issues such as these and new quality control tests are in development to automatically identify these images.

\begin{figure}
\centering
\includegraphics[width=0.48\textwidth]{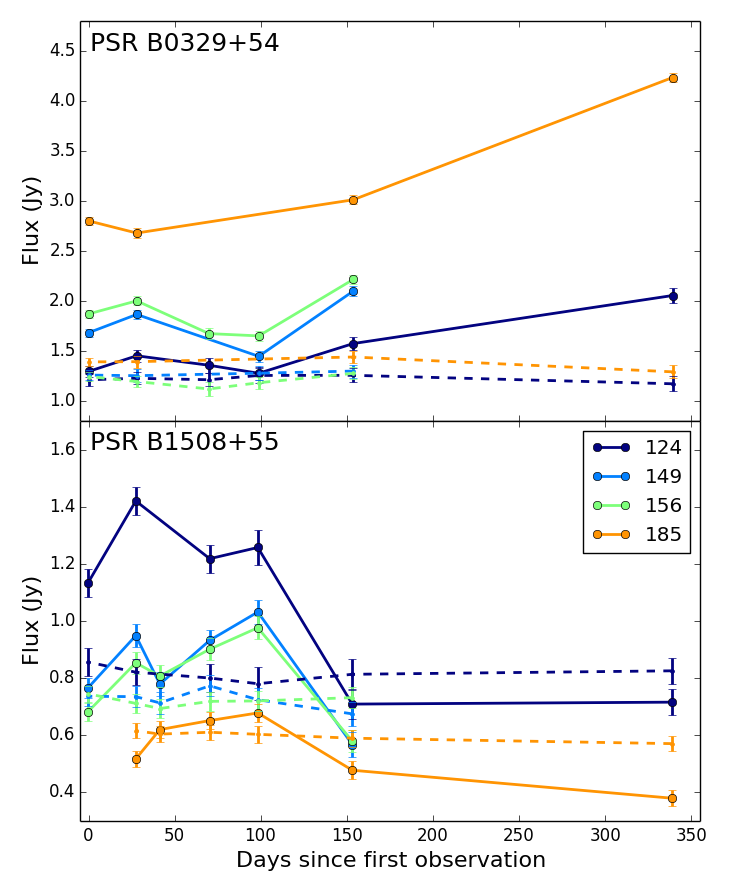}
\caption{The light curves of two scintillating pulsars, PSR B0329+54 (top) and PSR B1508+55 (bottom), identified using the anomaly detection strategy with a stable source in their respective fields shown for reference (dashed lines). The lines are colour coded by the observed frequency in MHz.}
\label{fig:pulsars}
\end{figure}

\begin{figure*}
\centering
\includegraphics[width=0.49\textwidth]{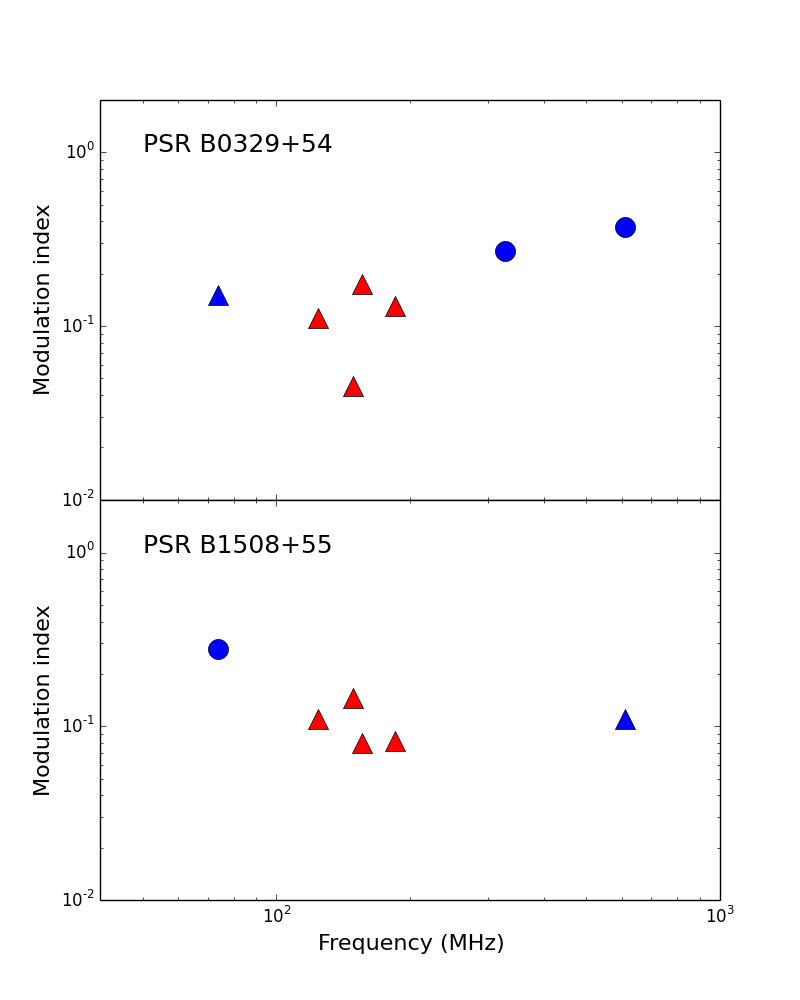}
\includegraphics[width=0.49\textwidth]{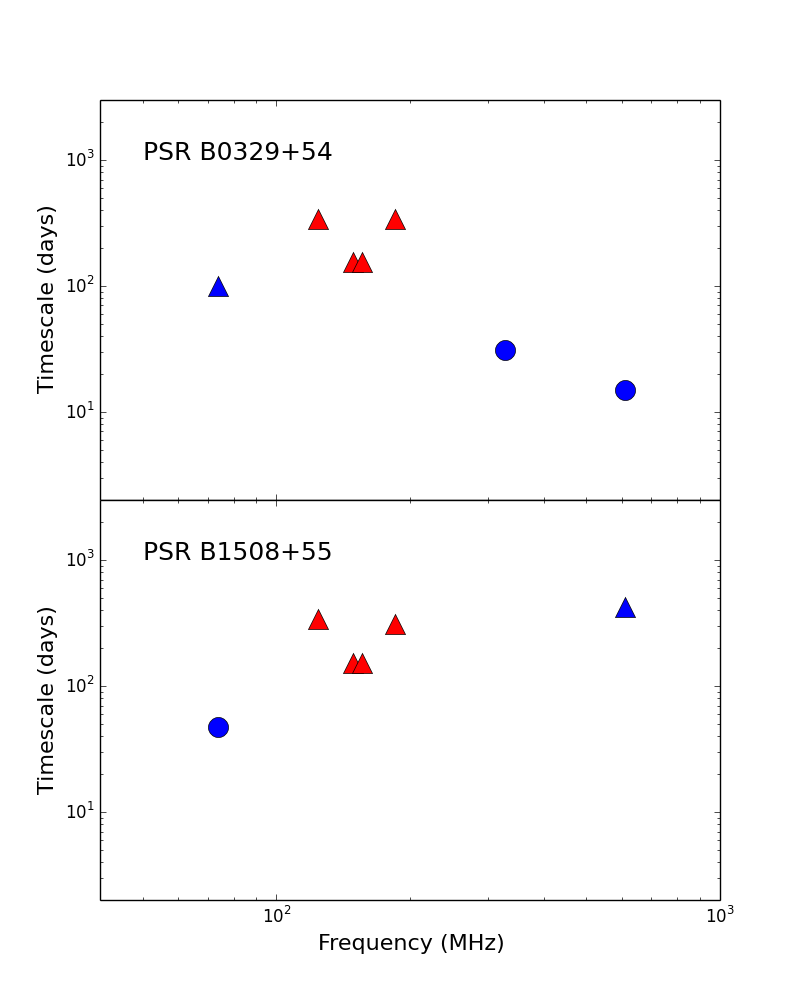}
\caption{The modulation index (left) and the modulation time scale (right) of the two scintillating pulsars, PSR B0329+54 (top) and PSR B1508+55 (bottom). The red triangles are the LOFAR RSM lower limits and the blue data points are the detections (circles) and lower limits (triangles) taken from \citet{Esamdin:2004} and references therein.}
\label{fig:pulsars3}
\end{figure*}

Two of the blindly detected variable source candidates were identified as real variable sources as other sources in the field had stable flux densities. These two sources are associated with pulsars PSR B0329+54 ($\eta_{\rm (185~MHz)}=252$ and $V_{\rm (185~MHz)}=0.225$) and PSR B1508+55 ($\eta_{\rm (124~MHz)}=35.4$ and $V_{\rm (124~MHz)}=0.276$), which are both known to scintillate at low frequencies \citep{Kaspi:1992,Gupta:1993,Stinbring:1996,Stinebring:2000,Shishov:2003,Esamdin:2004}. In Figure \ref{fig:pulsars}, we show the light curves of these pulsars together with a nearby stable source. We note there are missing measurements for some snapshots as a few of the images were rejected by the automated quality control scripts. The time scale, of order several months, and magnitude of the variability of these pulsars are consistent with the refractive interstellar scintillation models predicted by \cite{Esamdin:2004}. We use the lightcurves from our 4 observing frequencies to calculate the pulsar modulation index, the characteristic change in {the flux density of a pulsar caused by  refractive scintillation in the interstellar medium}, using the method outlined in \cite{Esamdin:2004}. The refractive interstellar scintillation is characterised using the structure function of the flux density variability defined by
\begin{eqnarray}
D(\tau) = \frac{1}{\langle F \rangle ^2}\sum\limits_{j=1}^n \frac{(I_j - I_{j+\tau})}{W_\tau}, \label{eqn:structureFunction}
\end{eqnarray}
where $\tau$ is the duration between observations, $I_j$ is the flux density in observation $j$, $I_{j+\tau}$ is the flux density in observation $j+\tau$, $W_\tau$ is the number of days with a given lag time and $\langle F \rangle$ is the average flux density \citep{Esamdin:2004}. This structure function reaches a saturation value, where the variability is intrinsic to the source, on the characteristic variability time scale. 
In the RSM dataset, we have insufficient snapshots on long time scales to constrain the saturation value and, hence, we can only provide approximate lower limits on the modulation indices and variability time scales. In Table \ref{tab:pulsars} we provide the LOFAR constraints and in Figure \ref{fig:pulsars3} we plot these values with the results and predictions of \cite{Esamdin:2004}. Frequent monitoring will constrain these parameters further.

\begin{table}
\begin{tabular}{llll}
\hline
Pulsar & Frequency & Modulation & Time scale \\
       & (MHz)     &                  & (days) \\
\hline
B0329+54 & 124 & $>$0.111 & $>$339 \\ 
         & 149 & $>$0.045 & $>$154 \\
         & 156 & $>$0.173 & $>$154 \\
         & 185 & $>$0.130 & $>$339 \\
\hline
B1508+55 & 124 & $>$0.110 & $>$339 \\
         & 149 & $>$0.146 & $>$153 \\
         & 156 & $>$0.080 & $>$153 \\
         & 185 & $>$0.082 & $>$311 \\
\hline
\end{tabular}
\caption{The modulation index and modulation time scale for the two scintillating pulsars we identified.}
\label{tab:pulsars}
\end{table}

\subsubsection{Logistic Regression Results}

As with the anomaly detection strategy described in Section 4.3.1, a number of the sources in the RSM dataset are identified as false positive results which are candidate variables in the dataset. Of the 7 sources identified, 6 are due to the calibration errors described in Section 4.1.1 and the final candidate was the previously identified variable source PSR B0329+54. PSR 1508+55 was not identified using the logistic regression method, because logistic regression makes a more optimal separation between the stable and variable sources. We assumed that all the RSM sources are stable sources, although there are calibration uncertainties causing the variability parameters of some stable sources to be increased. As a result the logistic regression strategy will be biased towards higher variability parameters. The four variability parameters of PSR 1508+55 used in this analysis are comparable to those of sources suffering from calibration uncertainties and hence is only just inside the stable source region using the optimal classification attained by the logistic regression algorithm. With improved calibration across the dataset, this bias would be reduced and PSR B1508+55 is then likely to be correctly classified by this algorithm.

\subsubsection{Transient Candidates}

By choosing the strategy which maximises the F-Score and the recall (as described in Section 4.1.3), we look for candidate transient sources which should be detectable in the best (or worst) rms region of the best previously studied image. Using these thresholds for the best and worst rms regions ($34\sigma$ and $2\sigma$ respectively), we identify 5 false positive results from the RSM dataset which are candidate transient sources. 4 of these candidates are caused by poor calibration in one of the initial pointing directions. The final candidate transient is caused by a spatially extended source that had been resolved into different components by {\sc PySE}. No transient sources were identified as part of this analysis. As discussed in Section 2.3, the hybrid dataset may be biased against low level variability and therefore further work is required to confirm that there are no transients.

\section{Future areas for development}

This paper has presented a number of strategies to automatically search for transient and variable sources using outputs from automated pipelines such as {\sc TraP}. These strategies have been successfully applied to observed and simulated datasets. However, there are a number of different areas which warrant further investigation with significantly larger datasets and more complex simulations.

\subsection{Improved training datasets}

As described in Section 2.3, the hybrid dataset has provided a reasonably representative training dataset to develop machine learning algorithms, which have aided in the identification of transient and variable sources in the observed RSM dataset. However, there are a number of caveats with this hybrid dataset with the most significant issue being the bias against low level intrinsic variability in the observed dataset. There are two options that could be investigated for future work:

\begin{itemize}
    \item Improve simulation strategy to better account for the expected, but difficult to model, sources of uncertainty, then produce fully simulated stable and variable sources that are representative of the observed dataset.
    \item Manually classify a large observed dataset including a significant number of known variable and transient sources.
\end{itemize}

\subsection{Time variability of variability parameters}

\begin{figure}
\centering
\includegraphics[width=0.48\textwidth]{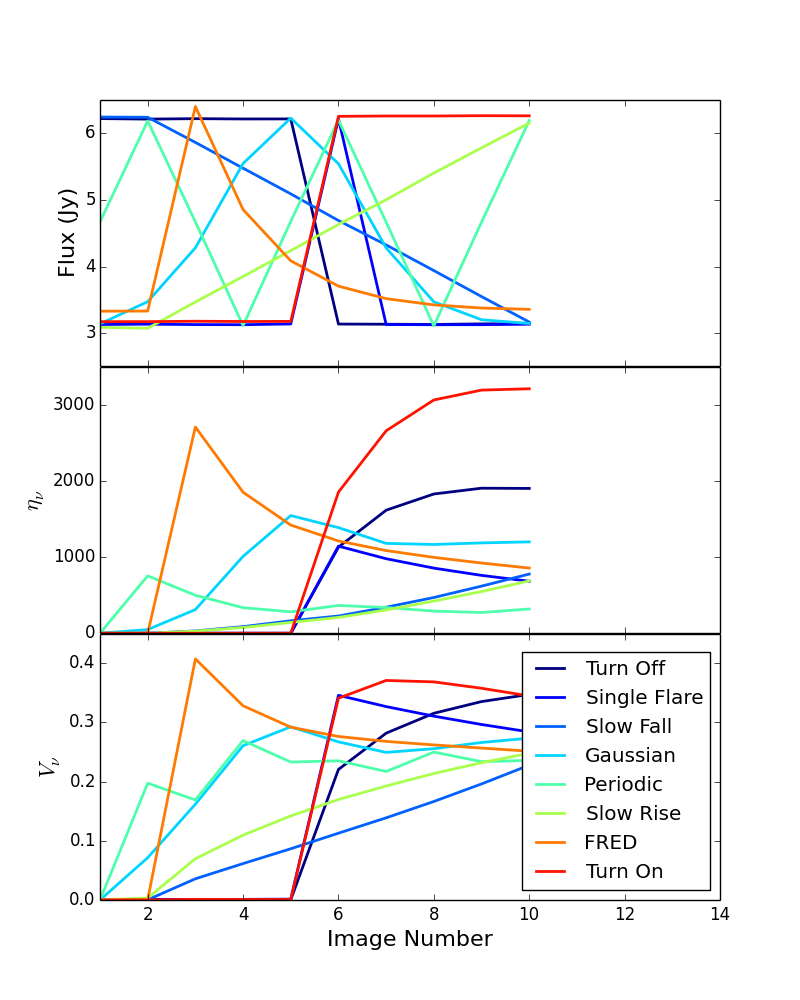}
\caption{The evolution of the variability parameters with time for each of the simulated variable sources with the same colour scheme as given in Figure \ref{fig:sim_trans_params2}. We show the flux density light curve (top), with the variability parameters $\eta_{\nu}$ (middle) and $V_{\nu}$ (bottom) evolution.}
\label{fig:timevariable}
\end{figure}

The variability parameters are all calculated using aggregate properties of the light curves; as new data points are added these parameters can change significantly. Hence a source which would be identified as variable after 10 images, may no longer appear variable after 100 images. For this reason, the {\sc TraP} records the variability parameters after each time step in individual light curves. In Figure \ref{fig:timevariable} we show the evolution of the variability parameters as a function of the light curve for the 8 different simulated source types presented in this paper. From this figure, it is clear that some sources have steadily increasing variability parameters; however a number of the source types have steadily decreasing parameters. Therefore, the variability can be missed if variable sources are only identified after processing a large number of images.

\begin{figure}
\centering
\includegraphics[width=0.5\textwidth]{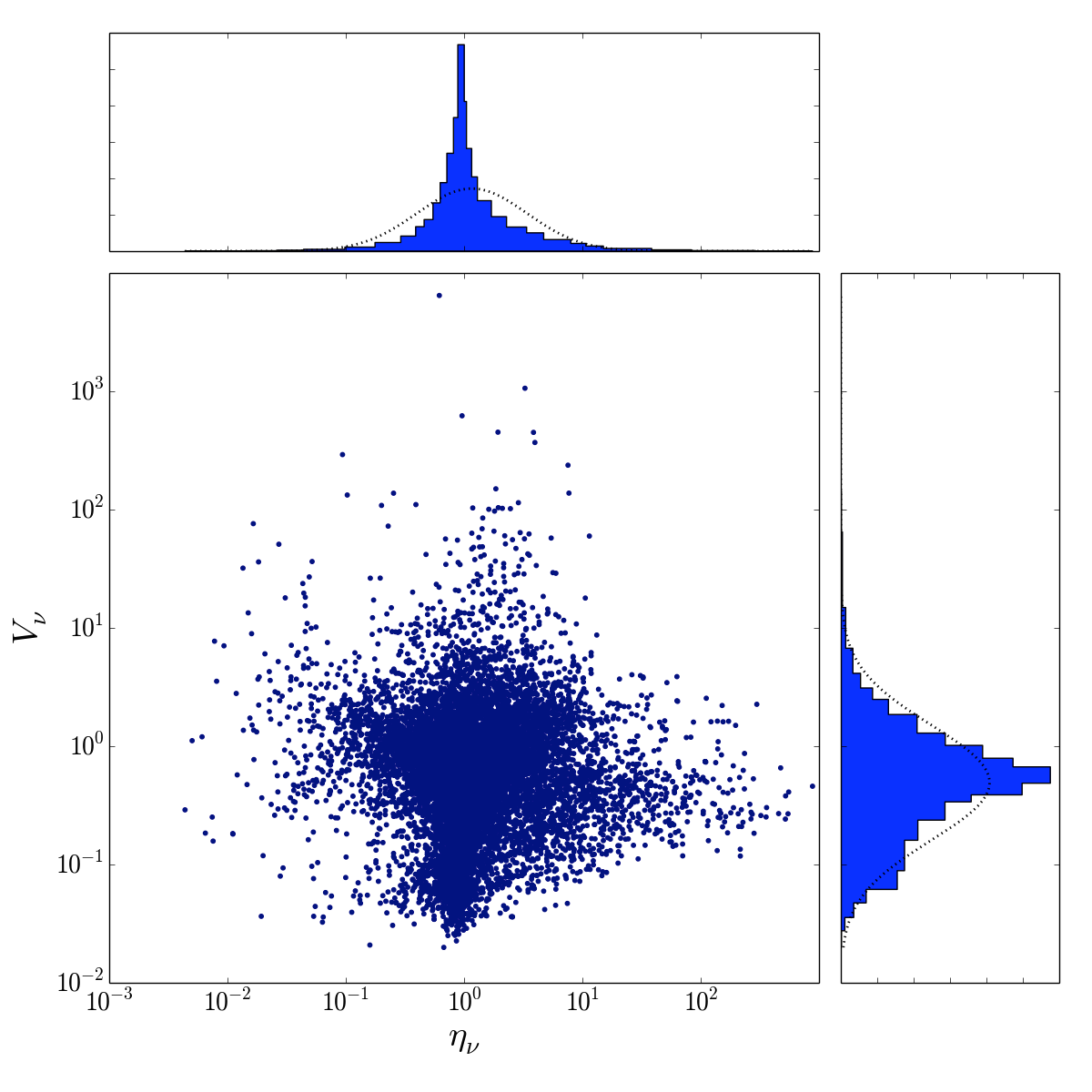}
\caption{The variability parameters for all the sources observed in the 3XMM dataset.}
\label{fig:rsm_trans_params1_5}
\end{figure}

Additionally, the variability parameters of the stable sources are also likely to stabilise to smaller values after large numbers of observations. Surveys with uneven sampling of fields are then unlikely to have stable sources which can be simply modelled as a single Gaussian distribution. For instance light curves from sources in the 3XMM dataset \citep{Watson:2009} may have anything from a few to thousands of data points. The variability parameters were calculated, using Equations \ref{eqn:eta_nu} and \ref{eqn:V_nu}, for the light curves of all sources in the 3XMM dataset and are shown in Figure \ref{fig:rsm_trans_params1_5}. They are clearly poorly fitted by a single Gaussian and is likely to be multiple Gaussian distributions superimposed on top of each other. This is most likely caused by the sources having different numbers of data points in their light curves and can be solved by comparing the light curves with the same number of data points. 
 
Both of these issues should be solved by simply applying the strategies presented in this paper at each time step or with increasing number of data points in the light curve. In the future, there will be significantly larger observed datasets and corresponding simulated datasets containing more snapshots. These larger datasets can be used to confirm that the variability parameters are dependent upon the number of data points in the source light curve. Additionally, using these larger datasets, the strategies outlined in this paper can be adapted into more complex algorithms that search for variability after each new data point is inserted into a light curve.

\subsection{Combining multi-frequency and polarisation information}

In this paper, we have focused on the variability parameters for the simulated simple light curves. A number of surveys have multi-frequency and polarimetric data which can provide extremely valuable additional information. For instance transients and variable sources detected at multiple frequencies are more likely to be real, while spectra, polarisation and variability delays between frequencies can give clues about the progenitors and emission mechanisms.

The logistic regression strategy presented in Section 4.2 has been developed to be easily adaptable to incorporate multiple new parameters. However, this strategy requires large training datasets containing both stable and simulated sources. Although the stable sources are easily obtained from existing surveys, there are still only a few transient and variable sources identified at low radio frequencies. Therefore we are still reliant on simulations and it is difficult to predict the multi-frequency and polarisation behaviour of the, as yet unknown, population of variable sources. Once significantly larger datasets have been obtained and there are more identified transient and variable sources, we will be able to develop these strategies.

\subsection{Supervised Machine Learning Strategies}

As the number of training parameters and the datasets increase, via the developments such as those suggested in Sections 5.1 -- 5.3, other supervised machine learning algorithms may perform better than the algorithms presented here and will require further investigation. These improvements may be in both the training speed and the classification accuracies. Other supervised algorithms that will become interesting future alternatives include:

\begin{itemize}
    \item Random Forests: A supervised machine learning algorithm that builds ensembles of decision trees for classification or regression \citep[e.g.][]{Breiman:2001}. These algorithms are at their best for studies where the categories well are known, with training sets containing sufficient sources in each category, and have been applied to a wide range of astronomical datasets \citep[e.g.][]{Lo:2014a,Lo:2014b,Farrell:2014}. At low frequencies, we do not have a well constrained population of sources to be able to produce the training set. This algorithm has the potential to be extremely useful in the future when the training dataset is of sufficient quality.
    \item Support Vector Machines: These work by identifying optimal boundaries known as hyper-planes between data points in multiple dimensions, similar to the linear divisions used by the logistic regression algorithm in this analysis. The data are first transformed into different dimensional space where they can be simply separated in a similar way to logistic regression \citep[e.g.][]{Cortes:1995}. These support vector machine algorithms are typically optimised for datasets with many training parameters. As this paper focuses on a relatively small dataset with few training parameters which can be linearly separated without transformation, we decided it was not necessary to use this strategy. As datasets become larger and more complex, support vector machines may become more appropriate.
\end{itemize}

\subsection{Unsupervised Machine Learning Strategies}

We have relied upon simulated sources to model the transient and variable source populations and utilised supervised machine learning strategies that are trained using the input datasets. When introducing the anomaly detection strategy, we noted this is typically an unsupervised machine learning strategy that can simply be applied to the dataset but relies upon choosing an appropriate $\sigma$ threshold. However, there are a number of other unsupervised machine learning strategies that are good at identifying clusters of sources from unlabelled datasets, such as k-means clustering algorithms \citep{Lloyd:1982}. These offer extremely promising strategies that will be able to identify unusual sources without assumptions about the properties of the source populations. Additionally, the unsupervised strategies do not require the large labelled training sets that can be challenging to obtain and are likely to contain biases against particular types of variability (see Sections 2.3 and 5.1). These strategies perform best with datasets with sufficient sources of each type. For our purposes, this equates to larger datasets, which will become available from various transient surveys running over the coming years, and warrant further investigation.

\section{Conclusions}

Reliable automated transient and variable source detection is essential with the expected influx of data from large transient surveys. Tools such as {\sc TraP} are essential for processing these datasets but require the astronomer to choose appropriate thresholds. This paper has presented methods for choosing these thresholds using machine learning strategies applied to observed and simulated datasets.

Using an anomaly detection strategy, adapted to be a supervised machine learning strategy in 2 dimensions, we have developed a method for astronomers to choose thresholds to meet specified criteria on the precision and recall of the classification of variable sources. This enables flexibility in the choice of thresholds; for instance the precision can be increased when requiring extremely reliable triggers to carefully control choice of follow-up targets. Alternatively, the recall can be set to higher values when the user wants to maximise the number of candidate variable sources. With reasonable input values for precision and recall, we blindly detected two known scintillating pulsars in the LOFAR RSM dataset. However, although the anomaly detection strategy performs better than the arbitrary thresholds chosen in previous studies, we note that this strategy is slow to train and provides sub-optimal classifications in comparison to the other methods presented in this paper.

The second strategy presented uses a logistic regression strategy that divides datasets into two populations in a multiple parameter space. We have used 4 parameters ($\eta_{\nu}$, $V_{\nu}$, maximum flux density and the ratio between the maximum flux density and average flux) but the algorithm can be easily adapted to include many more parameters. This method achieves a precision of 98\% and recall of 91\% with the dataset presented in this paper. Further analysis with larger datasets and a wider range of variable sources will be highly beneficial in further testing and extending this strategy.

Although transient sources will often have high values for their variability parameters, this is not always the case as their light curves may have insufficient data points to confirm variability or, after they have appeared, the sources may have a reasonably stable flux. {\sc TraP} has been specifically designed to deal with these sources by checking if they would be detected in the previous best images. To enable fine tuning, a margin can be applied to the detection threshold and we have demonstrated strategies to determine the optimal margin for the dataset.

These strategies have been developed using the {\sc TraP} and LOFAR observations. We note that the performance of the machine learning algorithms needs to be tested with significantly larger and varied datasets as they become available over the coming years. Moreover, these methods have been designed to be adaptable to other pipelines and wavelengths. In future work, we plan to extend these strategies and apply them to a wide range of transient and variable sources.

\section*{Acknowledgements}

The authors thank the anonymous referees for all of their comments as they helped improve this publication.

The authors would like to thank the ASTRON Radio Observatory, particularly Michiel Brentjens and Carmen Toribio, for their assistance in obtaining and processing the data used in this work. This work is supported in part by European Research Council Advanced Investigator Grant 247295, European Research Council Advanced Grant 267697 and Consolidator Grant 617199.

The Low Frequency Array designed and constructed by ASTRON, has facilities in several countries, that are owned by various parties (each with their own funding sources), and that are collectively operated by the International LOFAR Telescope (ILT) foundation under a joint scientific policy.

This work utilises a number of {\sc Python} libraries, including the {\sc Matplotlib} plotting libraries \citep{Hunter:2007} and {\sc astroML} \citep{Vanderplas:2012}. We also use the on-line code repository GitHub (https://github.com). Finally, the authors acknowledge training in machine learning strategies received from the Stanford Machine Learning course available on Coursera \\ (https://www.coursera.org/course/ml).

\section*{References}

\bibliographystyle{elsarticle-harv}
\bibliography{trap}

\end{document}